\documentclass{elsarticle} 

\usepackage[utf8]{inputenc}
\usepackage{geometry}
\usepackage{graphicx} 
\usepackage{natbib}
\usepackage{fleqn}
\usepackage{booktabs} 
\usepackage{array} 
\usepackage{paralist} 
\usepackage{verbatim} 
\usepackage{mathtools}
\usepackage{amsmath,amssymb,amsthm}
\usepackage{mathrsfs}
\usepackage{hyperref}
\usepackage{cleveref}
\usepackage{slashed}
\usepackage[justification=centering]{caption}
\usepackage{subcaption}
\usepackage{bbm}

\newcolumntype{C}[1]{>{\centering\let\newline\\\arraybackslash\hspace{0pt}}m{#1}}

\newcommand{\mc}[1]{\mathcal{#1}}
\newcommand{\ii}{\text{i}}

\newcommand{\dd}{\text{d}}
\newcommand{\db}{\bar{\partial}}

\newcommand{\ul}[1]{\underline{#1}}

\swapnumbers
\newtheorem{theorem}{Theorem}[subsection]
\newtheorem*{theorem*}{Theorem}
\newtheorem{prop}[theorem]{Proposition}

\newtheorem*{conj*}{Conjecture}

\newtheorem*{prin*}{Principle}

\theoremstyle{definition}

\newtheorem*{warning*}{Warning}

\begin{document}

\begin{frontmatter}

\title{On the geometry of magnetic Skyrmions on thin films}
\author{Edward Walton}
\ead{e.walton@damtp.cam.ac.uk}
\address{Department of Applied Mathematics and Theoretical Physics, University of Cambridge, Wilberforce Road, Cambridge, CB3 0WA, UK}

\begin{abstract}
We study the recently introduced `critically coupled' model of magnetic Skyrmions, generalising it to thin films with curved geometry. The model feels keenly the extrinsic geometry of the film in three-dimensional space. We find exact Skyrmion solutions on spherical, conical and cylindrical thin films. Axially symmetric solutions on cylindrical films are described by kinks tunnelling between `vacua'. For the model defined on general compact thin films, we prove the existence of energy minimising multi-Skyrmion solutions and construct the (resolved) moduli space of these solutions.
\end{abstract}

\end{frontmatter}

\section{Introduction}

Finding effective methods of storing data is one of the great challenges of the modern era. The aim of the game is to store ones and zeroes in a way that is stable against unwanted fluctuations and yet easily writable and readable. Importantly, one wants to store these ones and zeroes as densely as possible.

Magnetic Skyrmions have emerged as a potential player in high-density `next-generation' storage devices. They are topological solitons in two-dimensional magnetic materials (realised, for example, as thin films or as interfaces), characterised by a non-trivial winding of the magnetisation field (which is a three-dimensional vector field of approximately constant positive magnitude). They were predicted theoretically in \cite{polyakovMS, bogdanovMVS} and have been observed in chiral magnets (see, for example, \cite{muhlbauerMS,yuSC}). By virtue of their topological characterisation, they are stable (at least in the theoretical limit of genuinely two-dimensional materials). They can be manipulated, and they can be very small. 



Recently, there has been interest in models of chiral magnets admitting Skyrmions and other solitons of BPS (Bogomolny--Prasad--Sommerfield) type \cite{bartonMS, schroersGSM, hongoICM}. These are models admitting a class of classical solutions which obey a first order equation and which absolutely minimise the energy within a topological sector. This latter condition implies that constituent solitons in a multi-soliton BPS configuration (if they can be identified) exert no net force on each other. There is then a \emph{moduli space} of degenerate solutions to the first order equation, parameterised by the positions and `internal' moduli of the solitons. The geometry of this moduli space can give a good deal of information about the low energy (quantum) dynamics of the system. BPS models are very special, requiring fine-tuned values of the parameters of the model, and so are not always physically relevant, but they provide a wonderful theoretical testing ground for ideas. BPS models are ubiquitous in high energy physics (where the fine-tuning of parameters is sometimes justified by supersymmetry) and one can use the wealth of tools developed in that field in these condensed matter applications.

There is also mathematical and physical interest in Skyrmions on curved thin films (see \cite{gaidideiCE, streubelMCG}, for example). By manipulating the geometry of the film, one may be able to manipulate the Skyrmions -- it has been demonstrated that by creating a curvilinear defect, one can pin Skyrmions in place, at least within a certain family of micromagnetic models \cite{kravchukSCD}. Ideas like this hint at interesting mathematical links between the extrinsic geometry of thin films and Skyrmion solutions.

In  \cite{bartonMS},  a BPS model of magnetic Skyrmions on the plane was constructed. In \autoref{sec:mscc}, we generalise this model to curved thin films, possibly with interesting topology. The model has a Dzyaloshinskii--Moriya (DM) antisymmetric exchange interaction\footnote{This is in contrast to some other models of Skyrmions on curved thin films, which generate the DM interaction as a kind of fictional force using the geometry (see \cite{gaidideiCE}, for example).}. As in the models studied in \cite{gaidideiCE}, the geometry induces an apparent magnetic field normal to the film in our model. We write down the first order BPS equation for energy minimising solutions in the theory and prove the existence of solutions on general compact films (it was shown directly in \cite{schroersGSM} that local solutions can always be constructed; our global proof follows a simple and well-known exercise in complex algebraic geometry). We interpret solutions to this equation as describing Skyrmion states, with number of Skyrmions \(N\) given by the topological degree of the solution (as usual). As is typical for BPS models, the energy of solutions to the BPS equation is linear in \(N\). In general, Skyrmion and anti-Skyrmion density in the model is trapped in regions where the film has (extrinsic) curvature.

We go on in \autoref{sec:exact} to write down explicit solutions on certain symmetric films. In particular, we solve generally the BPS equation for Skyrmions on a round spherical film for every Skyrmion number. Among the solutions, we find the `hedgehog' solution. We also study the case of axially symmetric Skyrmions on certain axially symmetric films and find exact solutions on cones, which we view as a model for solutions on films with bumps. We further solve the theory on the cylinder. In this case, there are two axially symmetric vacuum solutions, and non-trivial axially symmetric Skyrmion solutions are kinks mediating between these vacua.

In \autoref{sec:moduli}, we consider the moduli space of solutions on general compact films. This is a singular complex manifold of complex dimension \(2N+1-g\), where \(N\) is the number of Skyrmions, and \(g\) is the genus of the film. It admits a natural resolution given by a certain moduli space of semi-local vortices in a background gauge field, which we describe. We give a direct construction of the moduli space of these vortices (at least for sufficiently high vortex number) using a `dissolving vortex' limit. 

In \ref{sec:nlsm} we collect some basic results (possibly of general interest) regarding two-dimensional nonlinear sigma models with target a symplectic manifold in the presence of a background gauge field for a Hamiltonian group action on the target space. In particular, we observe that the BPS energy bound for models of this type is given by the pull back of the equivariant symplectic form on the target space evaluated on the funadamental class of the two-dimensional domain. This clarifies the geometrical meaning of the terms that appear in the context of magnetic Skyrmion models.

\section{Magnetic Skyrmions on curved thin films at critical coupling}
\label{sec:mscc}

\subsection{An effective description of chiral magnetism from torsional geometry}

In \cite{bartonMS}, it was realised that the theory of planar micromagnetic thin films with DM interaction could be captured by the theory of a two-dimensional sigma model with target space the sphere in the background of a particular \(SO(3)\) gauge field.

To understand why this is a worthwhile approach, we should recall first that, in the micromagnetic regime, the magnetisation field (divided by the saturation magnetisation) is a true unit vector field, \(m\), in three dimensions. Hence, it is acted on naturally by the covariant derivative associated to an orthogonal connection (the connection should be orthogonal so that it preserves the unit length property) on the tangent bundle of a Riemannian three-manifold (usually \(\mathbb{R}^3\) or a domain thereof). It becomes clear that chiral interactions can be effectively modelled by allowing the connection to have torsion. 

As a very simple illustration of this, one can approximate the magnetisation field in the helical phase of a chiral magnet in \(\mathbb{R}^3\) to be a solution to the first order equations
\begin{align*}
\partial_1 m^a &= -\epsilon_{1ab}m^b \\
\partial_2 m^a &= 0 \\
\partial_3 m^a &= 0 \text{,}
\end{align*}
where \(a,b=1,2,3\). This can be written
\[
\dd_\mc{A} m = 0
\]
where \(\dd_\mc{A}\) is the covariant derivative associated to the affine connection \(\mc{A}\) with components
\begin{equation}
\label{eq:helex}
\mc{A}_{ab} \equiv \epsilon_{1ab}\dd x^1 \text{.}
\end{equation}
in Cartesian coordinates. (Here and throughout we will use \(\equiv\) to denote equality modulo gauge transformations -- that is, local coordinate transformations.)

In general, an orthogonal connection \(\mc{A}\) on the tangent bundle of a Riemannian manifold \(M\) can be characterised by its \emph{contorsion tensor}
\[
K \coloneqq \mc{A} - \Gamma \in \Omega^1(M,\mathfrak{so}(3))
\]
where \(\Gamma\) is the Levi-Civita connection (the unique torsion-free orthogonal connection). In Cartesian coordinates on flat \(\mathbb{R}^3\), the components of the Levi-Civita connection vanish. In the case of the connection defined in \eqref{eq:helex}, the contorsion tensor in the given Cartesian coordinates is simply \(K_{ab} \equiv \epsilon_{1ab}\dd x^1\). In what follows, this contorsion tensor will determine the antisymmetric exchange interaction of the theory.

Another natural gauge covariant object that one can associate to a connection \(\mc{A}\) is its curvature. In general, the curvature of \(\mc{A}\) can be given in terms of \(K\) as
\begin{align*}
F(\mc{A}) &= F(\Gamma + K) \\
	&= F(\Gamma) + \dd_\Gamma K + \frac{1}{2} [K, K]\text{.}
\end{align*}

To see how the DM interaction is related to the choice of \(K\) we consider the following. In three dimensions, the micromagnetic energy density functional should take the form
\[
\mc{E}_\mc{A} [m] = |\dd_\mc{A} m |^2 + V(m)
\]
where \(\dd_\mc{A} \) is the covariant derivative and \(V\) is a choice of potential. Now, writing Greek indices for vector indices and Latin indices for matrix indices for the adjoint representation of \(\mathfrak{so}(3)\) \footnote{In three dimensions, these indices should be identified, but it is useful to keep them separate in the name of maintaining sanity when discussing the two-dimensional theory.}, we have
\begin{equation}
\label{eq:kin}
|\dd_\mc{A} m |^2 = | \dd m|^2 + 2{\mc{A}_\mu}^a_b m^b \partial^\mu m_a + {\mc{A}_\mu}^a_b m^b \mc{A}^\mu _{ac}m^c\text{.}
\end{equation}

The story makes sense for arbitrary three-manifolds, but from now on we restrict ourselves to the (physically relevant) case that the three-manifold is \(\mathbb{R}^3\) with its standard flat metric. In this case, we may choose a gauge (given by Cartesian coordinates) in which \(\mc{A} \equiv K\). Thus, in these coordinates, the contorsion tensor is precisely the tensor, sometimes called the \emph{DM vector},  defining the DM term, which is the second term on the right-hand side of \eqref{eq:kin}. (The third term on the right-hand side of \eqref{eq:kin} contributes to the effective potential energy of the theory.) It is natural therefore to choose \(K\) to be translationally invariant, meaning that \(\dd_\Gamma K = 0\). In Cartesian coordinates \(\{x^\mu\}_{\mu = 1,2,3}\), we choose
\[
K \equiv K_\mu \dd x^\mu
\]
where \(K_\mu\) are constant elements of \(\mathfrak{so}(3)\). 
For such choices, the fact that \(F(\Gamma) = 0 \) and \( \dd_\Gamma K = 0\) implies that
\[
F(\mc{A}) = \frac{1}{2}[K,K] \text{,}
\]
which is also translationally invariant.

We might further restrict to connections of the form
\begin{equation}
\label{eq:invcon}
\mc{A} \equiv \kappa O^A_\mu T_A \dd x^\mu
\end{equation}
where \(\kappa\) is a real number, \(O\) is an orthogonal matrix of determinant 1, and the \(T_A\) form the fixed basis for the adjoint representation of \(\mathfrak{so}(3)\) given by
\[
(T_A)_{ab} = \epsilon_{Aab}
\]
(of course, different choices can be made by a redefinition of \(O\)).

These connections are characterised more invariantly by their parallel transport. Given any straight line \(l \subset \mathbb{R}^3\), the parallel transport along \(l\) is given by fixed rate rotations (with frequency given by \(\kappa\)) in a plane perpendicular to the line defined by acting on \(l\) with \(O\). The rate of rotation \(\kappa\) introduces a length scale into the theory. 

This shows us that, in general, these connections are symmetric with respect to the subgroup \(U(1)\) of the group \(SO(3)\) of global spatial rotations consisting of those rotations in the plane preserved by \(O\).

If one takes the special case \(O = 1\), one obtains the connection known as the \emph{Cartan spiral staircase}. This leads to the usual bulk DM term, of the form \(\kappa \textbf{m} \cdot \nabla \times \textbf{m}\), and will be the case we study primarily. It is the most symmetric choice, enjoying full rotational invariance, as \(O=1\) preserves every plane. It is for this reason that we focus on it: it treats every tangent plane of an arbitrary embedded two-dimensional surface in the same way. It will favour Bloch-type Skyrmions in two-dimensions. 

An interesting different choice is  
\begin{equation}
\label{eq:diffcon}
O = \left(
\begin{matrix}
0 &1&0\\
-1&0&0 \\
0&0&1 	
\end{matrix}
\right)
\end{equation}
which would give an `interfacial' DM term on reduction to the two-dimensional \((x_1,x_2)\) plane. This choice would favour N\'eel-type Skyrmions on that plane.

Now, if \(K_\mu = \kappa O^A_\mu T_A\), then
\begin{align*}
F(\mc{A})_{\mu\nu} &= \frac{1}{2} [K_\mu, K_\nu] \\
	&= \frac{\kappa^2}{2} \epsilon_{ABC}O_\mu^AO_\nu^BT_C \\
	&= \frac{\kappa^2}{2} \epsilon_{\mu \nu \lambda}  O^{\lambda C} T_C\text{.}
\end{align*}
Here we raise and lower Greek indices using the metric on \(\mathbb{R}^3\), which is just \(\delta_{\mu\nu}\) in these coordinates.
We then see that
\begin{align}
*F(\mc{A})_\mu &= \kappa^2 O_\mu^A T_A \nonumber\\
	&= 	\kappa K_\mu \text{.} \label{eq:curv}
\end{align}
This implies, after a similar computation, that
\begin{equation}
\label{eq:dcurv}
\dd_A^\dagger F(\mc{A}) = \kappa^2 K \text{,}
\end{equation}
so that \(\mc{A}\) is not a solution to the Yang--Mills equation, instead obeying a kind of non-Abelian Proca equation. 

We also observe that, for \(\kappa \neq 0\), there is no adjoint Higgs field \(\Phi\) such that \(F(\mc{A}) = *\dd_\mc{A} \Phi\), which is to say that \(\mc{A}\) does not form part of a BPS monopole solution. This follows because the BPS equation implies that 
\[
\dd_\mc{A} ^\dagger F(\mc{A}) = *[F(\mc{A}), \Phi]
\]
which gives, using \eqref{eq:curv} and \eqref{eq:dcurv},
\[
\kappa^2 K = \kappa [K, \Phi]
\]
which can't be solved componentwise in \(\mathfrak{so}(3)\) for \(\kappa \neq 0\).

Let us note briefly that while it is usually natural to impose translation invariance, rotationally covariant choices of connection are not the only interesting ones. Indeed, magnetic materials are often not isotropic in this way. An interesting example of this was studied recently in \cite{hongoICM}. 

\subsection{The critical micromagnetic energy functional on curved thin films}

Let us consider a thin film in \(\mathbb{R}^3\), which we idealise as a smoothly embedded two-dimensional surface
\[
i : \Sigma \hookrightarrow \mathbb{R}^3\text{.}
\]
The Euclidean metric \(g\) on \(\mathbb{R}^3\) induces a metric \(g_\Sigma \coloneqq i^*g\) on \(\Sigma\). In two dimensions, a metric induces a complex structure, so we may regard \(\Sigma\) as a Riemann surface with complex structure \(j_\Sigma\) and compatible metric \(g_\Sigma\). We write \(\textbf{N} = N^\mu \partial_\mu\) for the unit normal vector field to \(\Sigma \subset \mathbb{R}^3\), and
\begin{equation}
\label{eq:gaussmap}
n = \frac{N^1 + \ii N^2}{1+N^3} : \Sigma \to \mathbb{C}P^1
\end{equation}
for the corresponding Gauss map.

The metric connection \(\mc{A}\) on \(T\mathbb{R}^3\) induces an orthogonal connection 
\[
A \coloneqq i^*\mc{A}
\]
on the pullback bundle \(i^*T\mathbb{R}^3 \to \Sigma\), the restriction of the tangent bundle of \(\mathbb{R}^3\) to \(\Sigma\). This is a topologically trivial vector bundle of rank 3 on \(\Sigma\). We refer to \cite{schroersGSM} and to \ref{sec:nlsm} for comments on the model for a general connection. We will specialise from now on to the case that \(\mc{A}\) is the spiral staircase connection, so that
\[
A \equiv \kappa T_\mu \dd i^\mu
\]
where \(i^\mu = i^*x^\mu\) are the components of the map \(i\).

We consider the natural energy functional for the sigma model
\begin{equation}
\label{eq:energy}
E_A[m] = \frac{1}{2}\int_\Sigma *\left(|\dd_A m |^2 + V(m)\right)
\end{equation}
where \(V\) is a local potential energy function, \(*\) is the Hodge star on \((\Sigma, g_\Sigma)\) and \(|\cdot |^2\) denotes the square norm given by the combination of \(g_\Sigma^{-1}\) and the dot product of vectors. To relate \eqref{eq:energy} to a two-dimensional micromagnetic energy functional, so as to understand the role of the connection \(A\), we expand
\begin{align}
|\dd_A m|^2 &= |\dd m + A(m) |^2  \nonumber \\
	&= | \dd m |^2 + 2(A(m), \dd m) + | A(m)|^2 \label{eq:enex}
\end{align}
where the brackets \((\cdot, \cdot)\) denote the metric induced by the combination of \(g_\Sigma^{-1}\) and the dot product of vectors. The meaning of the first term on the right-hand side of \eqref{eq:enex} is clear: it is the usual symmetric exchange interaction. What about the other terms?

Being of zeroth order, the third term on the right-hand side of \eqref{eq:enex} contributes to the effective potential energy of the theory. To see what it is, it suffices to work locally on \(\Sigma\). Let \((x,y)\) be  coordinates on a local patch \(U \subset \Sigma\) such that the embedding \(i\) takes the local form
\begin{equation}
\label{eq:localembed}
i(x,y) = (x,y,f(x,y))
\end{equation}
for \(f:U \to \mathbb{R} \) a smooth real function (for sufficiently small \(U\), such local coordinates always exist, possibly after rotating the target space). The metric \(g_\Sigma\) takes the local form, writing \(\partial_x f = f_x\) and \(\partial_y f = f_y\),
\[
g_\Sigma |_U = (1+f_x^2)\dd x^2 + 2 f_x f_y \dd x \dd y + (1+f_y^2)\dd y^2 
\] 
(as usual, the juxtaposition of basis 1-forms denotes the symmetric product) so that the inverse is
\[
g_\Sigma^{-1} |_U = \frac{1}{1+f_x^2 + f_y^2} \left( (1+f_y^2) \partial_x^2 -2f_x f_y \partial_x \partial_y +(1+f_x^2)\partial_y^2\right)\text{.}
\]
The unit normal vector to \(i[U]\) in \(\mathbb{R}^3\) has the form
\begin{equation}
 \textbf{N} = N^\mu \partial_\mu = \frac{\partial_3 - f_x \partial_1 - f_y \partial_2}{\sqrt{1+f_x^2+f_y^2}}\text{.} \label{eq:normal}
\end{equation}
We also have
\begin{align*}
A|_U &= \kappa T_\mu \dd i^\mu \\
&= (T_1 + f_x T_3)\dd x + (T_2 +f_y T_3) \dd y  \text{.}
\end{align*}
Direct computation then reveals that, on \(U\),
\begin{align}
|A(m) |^2 &= \kappa^2\left( 1 + \frac{1}{1+f_x^2 +f_y^2} \left( f_x^2 m_1^2 +f_y^2 m_2^2 + m_3^2 - 2f_x m_1m_3 - 2 f_y m_2 m_3 + 2 f_x f_y m_1 m_2 \right) \right) \nonumber\\
	&= \kappa^2 \left( 1 + \left(\frac{1}{\sqrt{1+f_x^2+f_y^2} } (m_3 - f_x m_1 -f_y m_2) \right)^2\right) \nonumber \\
	&= \kappa^2 \left( 1 + m_N^2 \right) \label{eq:potcon}
\end{align}
where \(m_N \coloneqq \textbf{m} \cdot \textbf{N}\) is the normal component of the magnetisation field. Thus, the connection \(A\) induces a term in the effective potential which favours the tangential alignment of the magnetisation (it is an \emph{easy plane} contribution).

A similar computation reveals that the second term on the right-hand side of \eqref{eq:enex} is the standard DM term, \(2\kappa \textbf{m} \cdot \nabla \times \textbf{m}\), where \(\nabla \times\) here denotes the tangential curl. 
%

As shown in \cite{bartonMS,schroersGSM} and recalled in \ref{sec:nlsm}, there is a very special choice of potential for which the theory admits a so-called `Bogomolny rearrangement'. The choice is 
\begin{align}
V_\text{crit}(m) &= -2 ( * F(A))^B m_B \nonumber\\
	&= -2 \kappa^2  m_N \text{,} \label{eq:cpot}
\end{align}
where we have identified the tangent spaces to \(M\) with \(\mathfrak{so}(3)\) to view \(m\) as an \(\mathfrak{so}(3)\)-valued object \(m^BT_B\). To verify the second equality it again suffices to work locally. Again, suppose that the embedding \(i\) takes the local form of \eqref{eq:localembed}, determined by a function \(f\). Then we have, using \eqref{eq:curv},
\begin{align*}
F(A) &= i^*F(\mc{A}) \\ 
	&= \frac{\kappa^2}{2} i^* ({\epsilon}_{A\mu\nu} T^A \dd x^\mu \wedge \dd x^\nu ) \\
	&= \kappa^2 (T_3 - f_x T_1 - f_y T_2) \dd x \wedge \dd y \\
	&= \kappa^2 N^\mu T_\mu \sqrt{1+f_x^2+f_y^2} \, \dd x \wedge \dd y \\
	&= {\kappa^2} N^\mu T_\mu \omega_\Sigma
\end{align*}
where
\[
 \omega_\Sigma = \sqrt{1+f_x^2+f_y^2} \, \dd x \wedge \dd y
\]
is the Riemannian volume form on \(\Sigma\) (the quantity under the square root is the determinant of the metric \(g_\Sigma\)). Thus, \(*F(A)^B = \kappa^2N^B\), so
\[
*(F(A))^B m_B = \kappa^2 m_N
\]
as claimed. This contribution to the potential resembles a Zeeman term for an applied magnetic field normal to the thin film.

The overall effective potential, combining \(V_\text{crit}\) and the zeroth order contribution  \eqref{eq:potcon} from \(|\dd_A m|^2\), is 
\[
\kappa^2(1+m_N^2) - 2 \kappa^2 m_N = \kappa^2 (1-m_N)^2\text{.}
\]
This favours the alignment of the magnetisation with the normal vector to the surface. It is sensitive to the orientation of the surface.

The critically coupled energy functional can therefore be written as 
\begin{align}
E_\kappa (m) &= \frac{1}{2} \int_\Sigma * \left( |\dd_A m |^2 + V_\text{crit}(m) \right) \nonumber\\
&= \frac{1}{2}\int_\Sigma * \left( |\dd \textbf{m} |^2 + 2\kappa \textbf{m} \cdot \nabla \times \textbf{m} + \kappa^2 (1-m_N)^2 \right) \label{eq:mmef}
\end{align} 
where, as previously mentioned, the curl is the tangential curl. 

Some comments on \eqref{eq:mmef} can be made.
\begin{itemize}
	\item The energy functional \(E_\kappa\) is the natural generalisation of the functional of \cite{bartonMS} to curved films. In fact, in \cite{bartonMS}, a family of energy functionals on planar films was given, parameterised by \(U(1)\). Different members of this family can be produced by starting with different three-dimensional contorsion tensors which are symmetric under rotations in the plane parallel to the embedded planar film. The spiral staircase connection is one such choice, another is the example given in \eqref{eq:diffcon}. A general curved film has no global symmetry and so we do not naturally obtain a continuous family of models in our more general setting.
	
	\item The theory has a gauge symmetry, given by \(SO(3)\) gauge transformations of \(A\) and \(m\). From a three-dimensional perspective, this is related to invariance under local coordinate transformations, but in two dimensions we may view it as an abstract gauge symmetry. Both \(|\dd_Am|^2\) and \(V_\text{crit}(m)\) are separately invariant under these gauge transformations, although the decomposition of \eqref{eq:mmef} into a gradient energy, DM energy and effective potential energy is not invariant. Notice that, because \(A\) has curvature, it is not possible to set \(A\equiv 0\) and so remove the DM term using a smooth gauge transformation.
	
	\item It is not necessarily obvious that \(E_\kappa\) is bounded below, because the DM interaction term (and \(V_\text{crit}\)) can be negative. However, we show in \autoref{subsec:bog} that it is bounded below (on compact films) by a topological energy contribution.

	\item The energy functional \(E_\kappa\) depends on the single real parameter \(\kappa\), which introduces a length scale \(\frac{1}{\kappa}\) into the theory. If \(\kappa = 0\), then the model becomes the basic sigma model, which is conformally invariant and does not see the extrinsic geometry of the thin film. If one takes \(\kappa \to \infty\), then the potential energy dominates and one expects solutions to be normal to the film almost everywhere. In general, \(\frac{1}{\kappa}\) is the Skyrmion size, measuring the length scale over which a configuration deviates from the normal field.

	\item In the case that \(\kappa \neq 0\), the theory is highly sensitive to the extrinsic geometry of the thin film \(\Sigma \hookrightarrow \mathbb{R}^3\). This is unusual for soliton models, which often depend on intrinsic, but not extrinsic, geometry.
	
	This idea is clearly illustrated by consideration of the configuration \(\textbf{m} = \textbf{N}\). One might expect this to be a good `ground state' for the theory. After all, it has \(m_N=1\) and so minimises the potential energy \(\frac{\kappa^2}{2} \int_\Sigma* (1-m_N)^2\). However, its gradient energy density is generally non-zero, being given by
	\[
	\frac{1}{2} | \dd \textbf{N} |^2 = \frac{1}{2}(\rho_1^2 + \rho^2_2) =  2H^2 - G
	\]
where \(\rho_1, \rho_2\) are the principal curvatures of the embedded thin film, \(H = \frac{1}{2}(\rho_1+\rho_2)\) is the (extrinsic) mean curvature of the film, and \(G = \rho_1\rho_2\) is its (intrinsic) Gaussian curvature. Here we have used that \( \dd \textbf{N}\) is the \emph{shape operator} of the embedded surface \(\Sigma\) and has eigenvalues \(\rho_1, \rho_2\).

The DM energy density for the normal vector field is a total derivative and so the DM energy vanishes on compact surfaces. Then the overall energy of the configuration \(\textbf{m} = \textbf{N}\) on a compact Riemann surface \(\Sigma\) of genus \(g\) is
\begin{align}
E_\kappa (\textbf{N}) &= \int_\Sigma *\left(2H^2-G\right) \nonumber \\
	&= 2\int_\Sigma * ( H^2 ) + 4\pi(g-1) \nonumber \\
	&= 2\int_\Sigma *(H^2-G) + 4\pi (1-g) \label{eq:elastic}
\end{align}
where we have used the Gauss--Bonnet theorem to find the second equality, and we have added and subtracted \(2G\) and used the Gauss--Bonnet theorem to find the third equality. Notice that this does not depend on the parameter \(\kappa\). This result pre-empts part of the discussion of \autoref{subsec:bog} -- the Bogomolny argument given there might be viewed as a natural generalisation of these ideas about the curvature of embedded surfaces. In particular, we see that if \(H^2=G\), which is true for the round 2-sphere, then the Gauss map minimises the energy within its topological class (the degree of the Gauss map is \(1-g\)) and so it will give us a Skyrmion solution in the sense of \autoref{subsec:bog}.

Let us remark that the value of the energy functional \eqref{eq:elastic} for \(\textbf{N}\) gives rise to a functional on the space of smooth embeddings \( \Sigma \hookrightarrow \mathbb{R}^3 \) (which one may subject to some further constraints). Interpreted in this way, it is a natural energy functional which is well-studied in the context of elastic membranes (see \cite{hsuSMC}, for example). Allowing for noncompact surfaces, solutions to the corresponding Euler--Lagrange equations include minimal surfaces, which have \(H=0\).
\end{itemize}

\subsection{The Bogomolny equation and magnetic Skyrmions}
\label{subsec:bog}

The purpose of choosing the critical potential \(V_\text{crit}\) is, as shown in \cite{schroersGSM} and in \ref{sec:nlsm}, that one can rearrange the energy functional as follows (note that this rearrangement can be made for any choice of connection \(A\)). One has
\begin{equation}
\label{eq:en}
\frac{1}{2}\int_\Sigma * \left( |\dd_A m |^2 + V_\text{crit}(m) \right) = \int_\Sigma *|\db_A m |^2 + \int_\Sigma \left( m^*\omega_{S^2} + \dd (m \cdot A) \right) \text{,}
\end{equation}
where \(\omega_{S^2}\) is the standard symplectic form on the target 2-sphere with area \(4\pi\), and by \(m \cdot A\) we mean \(m_B A^B_\mu \dd x^\mu\).
Here, 
\[
\db_A m \coloneqq \frac{1}{2} \left( \dd_A m + J_{S^2} \circ \dd_A m \circ j_\Sigma \right)
\]
where \(j_\Sigma\) is the complex structure on \(\Sigma\) induced by its metric, and \(J_{S^2}\) is the standard complex structure on the target sphere. Locally, thinking of \(m\) as a unit three-vector \(\textbf{m}\), this has a component
\begin{equation}
\label{eq:vecbog}
(\db_A \textbf{m} )_1 = \frac{1}{2} \left( D_1 \textbf{m} + \textbf{m} \times D_2 \textbf{m} \right)
\end{equation}
where \(D_k\) is the \(k^\text{th}\) component of \(\dd_A\) in local conformal coordinates on \(\Sigma\), and the second local component of \(\db_A m \) is not independent.

Note that the second integral on the right-hand side of \eqref{eq:en} is topological (on a compact surface or with reasonable boundary conditions): it is the sum of \(4\pi\) times the Skyrmion number (the degree \(N\) of the map \(m\)) and an integrated `vorticity', which vanishes on closed surfaces. We give a more precise understanding of this topological energy, as a pairing in equivariant (co)homology, in \ref{sec:nlsm}. It has been argued in \cite{bartonMS} and elsewhere that the integrated vorticity contribution should be removed.

The rearrangement \eqref{eq:en} implies that the energy within a topological class is minimised for solutions of the first order \emph{Bogomolny equation}
\begin{equation}
\label{eq:bog}
\db_A m = 0 \text{.}
\end{equation}
The main aim of this note is to understand the solutions to this equation, which we abuse language to call \emph{(magnetic) Skyrmions}, and their moduli. The energy of solutions to this equation is \(4\pi N\), linear in the Skyrmion number \(N\), at least when the boundary term vanishes. This linear dependence of the energy on the topological degree is typical for solutions to Bogomolny equations in BPS models.

The cross product in \eqref{eq:vecbog} is often inconvenient to deal with, and so it is useful to change into coordinates in which \(J_{S^2}\) is simply \(\ii\). We do this by stereographic projection of the sphere to the extended complex plane, which carries a natural complex coordinate 
\[
v \coloneqq \frac{m_1 + \ii m_2}{1+m_3} \text{.}
\]
In this coordinate, the Bogomolny equation becomes
\begin{equation}
\label{eq:bogv}
\db v = - A^{0,1}(v)
\end{equation}
where \(A^{0,1}\) is the \((0,1)\) part (that is, the part proportional to \(\dd \bar{z}\) for \(z\) a complex conformal coordinate on \(\Sigma\)) of the connection \(A\). Of course one can not truly escape the nonlinearity inherent to equation \eqref{eq:bog}. Here it is tied into the action of \(A^{0,1}\) on \(v\), which is not a linear action: \(\mathfrak{so}(3)\) acts on the Riemann sphere by linearised (real) M\"obius transformations, or equivalently by (real) holomorphic vector fields. A particular advantage of this coordinate choice is that it allows us to use the tools of complex geometry, which we exploit in \autoref{subsec:proj}.

For now, let us compute the form of the Bogomolny equation in terms of the data of the embedding \(i: \Sigma \to \mathbb{R}^3\). In terms of our choice of connection \(A \equiv \kappa \left( \dd i^\mu \right) T_\mu\), the Bogomolny equation in the form of equation \eqref{eq:bogv} is
\[
\partial_{\bar{z}} v = - \kappa ( \partial_{\bar{z}} i^\mu ) T_\mu (v)
\]
where \(z\) is a (local) conformal complex coordinate on \(\Sigma\). Recall that a local complex coordinate \(z\) is \emph{conformal} if, locally,
\[
g_\Sigma = \Omega^2(z,\bar{z}) \dd z \dd \bar{z}
\]
for a real positive function \(\Omega^2\), the \emph{conformal factor}.

Once more, it is convenient to work in a local patch \(U \subset \Sigma\) with coordinates \((x,y)\) such that the embedding \(i\) takes the form of \eqref{eq:localembed}, determined by the single function \(f\).
Then the Bogomolny equation becomes
\begin{equation}
\label{eq:bogloc}
\partial_{\bar{z}} v = - \frac{1}{2} \kappa \left(\partial_{\bar{z}} (x+\ii y) (T_1 -\ii T_2)  + \partial_{\bar{z}}(x- \ii y) (T_1+\ii T_2) +2 ( \partial_{\bar{z}} f )T_3    \right) (v)\text{.}
\end{equation}
The obvious complex coordinate \(u \coloneqq x+\ii y\) on \(U\) is generally not conformal.

The question of how to produce a conformal coordinate \(z\) from a general coordinate \(u\) is answered by Gauss's theory of isothermal coordinates. It can be shown that a coordinate \(z(u,\bar{u})\) is conformal if it obeys the \emph{Beltrami equation}
\begin{equation}
\label{eq:bel}
\frac{\partial z}{\partial u} = \mu \frac{\partial z}{\partial \bar{u}}
\end{equation}
where (for our choice of embedding)
\[
\mu \coloneqq \frac{f_x^2 - f_y^2 + 2\ii f_x f_y}{2+f_x^2+f_y^2 +2 \sqrt{1+f_x^2+f_y^2}} 
\]
is the \emph{Beltrami coefficient} of the metric \(g_\Sigma\). One may note that
\[
\frac{f_x^2 - f_y^2 + 2\ii f_x f_y}{2+f_x^2+f_y^2 +2 \sqrt{1+f_x^2+f_y^2}} = \left(\frac{f_x +\ii f_y}{1+\sqrt{1+f_x^2+f_y^2}}\right)^2
\]
which reveals that \(\mu = n^2\), the square of the Gauss map \eqref{eq:gaussmap} (recall from \eqref{eq:normal} the form of the unit normal vector field).

Now let us get to grips with the pieces of \eqref{eq:bogloc}. Equation \eqref{eq:bel} implies that
\[
\partial_{\bar{z}} u = - \mu \partial_{\bar{z}} \bar{u}\text{.}
\]
Also,
\[
\partial_{\bar{z}} f = \bar{u}_{\bar{z}} \left( -\mu f_u + f_{\bar{u}} \right)\text{.}
\]
A computation then uncovers the fact that
\[
 -\mu f_u + f_{\bar{u}} = -n\text{.}
\]

The only piece of the puzzle left to find is the action of the \(T_A\) on the Riemann sphere coordinate \(v\). The action of \(\mathfrak{so}(3)\) is by (real) linearised M\"obius transformations, or equivalently, by the action of (real) holomorphic vector fields. One can show that
\begin{align*}
T_3 (v) &= -2 \ii v \\
(T_1 - \ii T_2) (v) &= 2\ii \\
(T_1 + \ii T_2) (v) &= - 2 \ii v^2 \text{.}
\end{align*}

We can now write down the Bogomolny equation in the simple form
\begin{align}
\partial_{\bar{z}} v &=  -\ii \kappa \bar{u}_{\bar{z}} \left( -\mu - v^2 - 2(-\mu f_u +f_{\bar{u}}) v \right) \nonumber \\
	 &=  -\ii \kappa \bar{u}_{\bar{z}} \left( -n^2 - v^2 + 2n v \right) \nonumber \\
	&= \ii \kappa  \bar{u}_{\bar{z}} \left( v-n\right)^2 \text{.} \label{eq:boggen}
\end{align}
In what follows, we will compute the geometrical prefactor \(\bar{u}_{\bar{z}}\) explicitly in examples. Note that, while we fixed the local form of \(u\), we did not fix \(z\), instead just asking that it solve the Beltrami equation \eqref{eq:bel}. One needs to pick a particular solution to compute \(\bar{u}_{\bar{z}}\) (this must be true, as the left-hand side of \eqref{eq:boggen} depends on the choice). However, we can say something general: direct computation reveals that
\[
| \bar{u}_{\bar{z}} | = \frac{\Omega(z,\bar{z})}{1+|n|^2} \text{,}
\]
where \(\Omega\) is the square root of the conformal factor \(\Omega^2\). 

As one would expect, this equation depends sharply on the extrinsic geometry of the thin film. It captures the interplay in the energy functional \eqref{eq:mmef} between the symmetric exchange, or gradient, energy \(|\dd m|^2\) and the potential energy \((1-m_N)^2\). If the Gauss map \(n\) is holomorphic, then there is a natural `ground state' \(v=n\). This clearly minimises the potential energy. In general, however, the Gauss map is not holomorphic, and the potential energy minimising configuration \(v=n\) has too much gradient energy to be an overall energy minimiser. On the other hand, a constant configuration (which minimises the gradient energy) generally has too much potential energy to be a solution.
	
	On a flat film the Gauss map is constant, and so trivially holomorphic. If one introduces a smooth bump into the film, then the constant solution is no longer an energy minimiser and so one expects the formation of some Skyrmion--anti-Skyrmion density in the bump. The idea that Skyrmion density might be pinned in place by bumps in a film is familiar from \cite{kravchukSCD}.
	
	Writing \(\tilde{v} = v-n\), the equation \eqref{eq:boggen} becomes
	\[
	\partial_{\bar{z}} \tilde{v} = \ii \kappa \bar{u}_{\bar{z}} \tilde{v}^2 + \partial_{\bar{z}} n
	\]
The quantity \(\partial_{\bar{z}} n\) is related linearly to the difference between the principal curvatures of the film.

\subsection{Projective bundles, their sections, and the existence of Skyrmions}
\label{subsec:proj}

We have seen that solutions to the Bogomolny equation are sections of a particular bundle over \(\Sigma\) with fibre \(S^2\) satisfying the Dolbeault-type equation \eqref{eq:bog}. We will address this structure from a complex analytic point of view. Namely, we regard these solutions as sections of a  projective line bundle (that is, a holomorphic fibre bundle with fibre \(\mathbb{C}P^1\)), where the holomorphic structure is determined by the operator \(\db_A\) (that \(\db_A\) defines an integrable holomorphic structure follows simply for dimensional reasons: there are no \((0,2)\)-forms on a Riemann surface). We can make some progress simply by recalling some basic, and well-known, facts about these objects.

A useful way to build projective line bundles is to take a rank 2 complex vector bundle and projectivise its fibres (that is, we replace the linear fibre by the space of its one-dimensional subspaces). If \(\mc{E}\) is a holomorphic vector bundle of rank 2, we write \(P(\mc{E})\) for the projective line bundle obtained by projectivising the fibres, which we call the \emph{projectivisation} of \(\mc{E}\).

At the level of transition functions, projectivisation corresponds to passing along the quotient map
\[
GL(2, \mathbb{C}) \to PGL(2, \mathbb{C}) \text{,}
\]
which has kernel \(\mathbb{C}^*\).
More precisely, the transition functions of \(\mc{E}\) live in the first (\v{C}ech) cohomology of the sheaf of holomorphic \(GL(2,\mathbb{C})\)-valued functions, while those of the projective bundle live in the first cohomology of the sheaf of holomorphic \(PGL(2,\mathbb{C})\)-valued functions. Taking the cohomology of the short exact sequence induced by the above map gives the long exact sequence
\[
\cdots \to H^1(\mc{O}^*) \to H^1(\mc{O}_{GL(2,\mathbb{C})}) \to H^1(\mc{O}_{PGL(2,\mathbb{C})}) \to H^2(\mc{O}^*) \to \cdots 
\]
where by \(\mc{O}_G\) we mean the sheaf of holomorphic \(G\)-valued functions for a group \(G\), and \(\mc{O}^*\) is the sheaf of holomorphic \(\mathbb{C}^*\)-valued functions.

We first notice that \(P(\mc{E} \otimes L) \cong P (\mc{E})\) for any line bundle \(L\) (this is simply the fact that multiplying the transition functions of \(\mc{E}\) by \(\mathbb{C}^*\)-valued transition functions does not affect the projectivisation). We further see that any obstruction to lifting a projective bundle \(P\) to a vector bundle \(V\) such that \(P = P(V)\) lives in \(H^2(\mc{O}^*)\).

In our case, \(\Sigma\) is compact and of real dimension two and therefore \(H^2(\mc{O}^*)= 0 \) (one can use the exponential exact sequence to see this). Hence, every projective line bundle on a Riemann surfaces arises as the projectivisation of some rank 2 vector bundle (in fact, a similar argument shows that every projective bundle on a compact Riemann surface arises as the projectivisation of a vector bundle).

We are interested in sections of projective line bundles. Now, holomorphic sections of the projectivisation \(P(\mc{E})\) of a vector bundle \(\mc{E}\) are equivalent to holomorphic line sub-bundles of \(\mc{E}\) -- a line in a fibre of \(\mc{E}\) is a point in the corresponding fibre of the  projectivised bundle. In one complex dimension, every rank 2 bundle has a holomorphic line sub-bundle (because Serre vanishing tells us that \(H^0(\mc{E}\otimes N) \neq 0\) for a line bundle \(N\) of sufficiently high degree) and so, independently of any details of the model at hand, we have the following basic existence result.

\begin{prop}
\label{prop:exist}
Solutions to the Bogomolny equation \eqref{eq:bog} on a compact surface always exist. More precisely, there exists an integer \(N_0\) such that there exist Skyrmion solutions of Skyrmion number \(N\geq N_0\).
\end{prop}

Note that this implies that solutions may exist with arbitrarily high degree - one can have arbitrarily large Skyrmion density. This is in contrast to vortices, which have some non-zero size and so have finite maximum density. We will see in \autoref{sec:moduli} that Skyrmions can be interpreted as particular two-flavour Abelian vortices in the limit that the vortex size goes to zero. Of course, from a physical perspective, the continuum approximation will break down at very high Skyrmion densities.

Notice that the proof of \cref{prop:exist} makes no reference to the choice of connection \(A\), which determines the holomorphic structure \(\db_A\). Thus, the result holds for any choice of DM vector (that is, any choice of three-dimensional contorsion tensor).

\subsection{Emergent electromagnetism}

From \eqref{eq:en}, the topological Skyrmion energy density is
\[
\mc{E}_\text{top}(m) = m^* \omega_{S^{2}} + \dd \left( m \cdot A \right) \text{.}
\]
This may be interpreted as an emergent magnetic field, generated by the magnetically charged Skyrmions. The corresponding emergent Abelian gauge potential is
\[
a = m^* \lambda_{S^2} + m \cdot A
\]
where \(\dd \lambda_{S^2} = \omega_{S^2} \), so that \(\lambda_{S^2}\) is the usual Dirac monopole connection on the target sphere. Under this interpretation, the Skyrmion density becomes the magnetic flux density. In this picture, Skyrmions are the two-dimensional shadow of `emergent' Dirac monopoles.

One can create or destroy Skyrmions on a thin film by passing an emergent monopole through the film. This follows simply from flux conservation. This provides a `low energy' realisation of the Skyrmion (un)winding via monopoles, as described in \cite{mildeUSL}.

A `UV-completion' of this theory (to allow for films with non-zero thickness) would presumably involve replacing these Dirac monopoles with BPS-type monopoles. It would be interesting to understand possible theories of this type, and in particular to study the expected pair creation of confined monopole-anti-monopole pairs  mediating the phase transition between the Skyrmion and helical phases \cite{mildeUSL,kanazawaCP}.

\section{Skyrmion solutions}
\label{sec:exact}

\subsection{Skyrmions on spherical films}

Solutions to the Bogomolny equation \eqref{eq:bog} on the flat plane were constructed in \cite{bartonMS}. The next simplest case to study is the round sphere.

The sphere is a good testing ground for these ideas due to its high degree of symmetry. Moreover, its Gauss map is holomorphic so, as we know from \eqref{eq:boggen}, the Bogomolny equation will separate and there is a good chance that we can find a general solution.

Solutions on the sphere are also interesting in that they may provide insight into Bloch points in three dimensions: If the Skyrmion number on the sphere is non-zero, any attempt to continue the solution into the interior of the sphere inevitably forces one to allow the magnetisation to go to zero at one or more points. We might therefore view these solutions on the sphere as the two-dimensional shadow of a smooth description of Bloch points viewed from large distances.

Let 
\[
i : S^2 \to \mathbb{R}^3
\]
be the standard round embedding of a sphere of radius \(R\). In terms of Cartesian coordinates \((x_1,x_2,x_3)\) on \(\mathbb{R}^3\), we introduce a complex coordinate on the sphere:
\[
z = \frac{x_1 + \ii x_2}{R+x_3} 
\]
which is 0 at the north pole and \(\infty\) at the south pole. The round sphere has the property that the Gauss map is the identity map, so we may write \(n = z\).

To write down the Bogomolny equation \eqref{eq:boggen} locally, we need to compute \(\bar{u}_{\bar{z}}\) where \(u =  x_1 +\ii x_2 \) (recall that, in the derivation of \eqref{eq:boggen}, the form of \(u\) was imposed from the start - there is no freedom in this definition). Computing the derivative reveals that
\[
\bar{u}_{\bar{z}} = \frac{2 R^2}{(R^2+|z|^2)^2}\text{.}
\]
This is the conformal factor of the induced metric on the sphere. That this is the case is special to the sphere. Then, the Bogomolny equation is 
\[
\partial_{\bar{z}} v = \frac{2\ii\kappa R^2}{(R^2+|z|^2)^2} \left( v - z \right)^2\text{.}
\]
This equation is separable and can be solved explicitly. The solution is
\[
v(z, \bar{z}) = z \left( 1 + \frac{R^2+|z|^2}{2 \ii \kappa R^2 - zf(z)(R^2 + |z|^2)} \right)\text{,}
\]
for any meromorphic function \(f\). Notice that if \(\kappa = 0\), one is left simply with a meromorphic function, as one should be. 

If we choose \(1/f = 0\), then we obtain the natural `ground state', \(v = z\). This is the `hedgehog' solution, where the magnetisation points radially. It is the Gauss map of the embedded sphere. While we call this the `ground state', it actually has degree 1, and so has positive energy \(4\pi\). If the radius is large, then, in a neighbourhood of the north pole, the Gauss map \(z\) appears constant and zero which is the natural ground state on the plane.

If \(f = 1/z\), we obtain
\begin{align}
v &= z \left( 1 + \frac{R^2 + |z|^2}{2\ii \kappa R^2 - (R^2 + |z|^2)} \right) \nonumber\\
	&= z \frac{2\ii \kappa R^2}{2\ii \kappa R^2 + (R^2 + |z|^2)} \text{.} \label{eq:dv}
\end{align}
This is a degree zero solution with zeroes at the north and south poles. As a degree zero solution, this configuration has zero energy and is therefore a vacuum of the theory.
\begin{figure}
\centering
\begin{subfigure}{0.4\textwidth}
\centering
\includegraphics[height = 5cm]{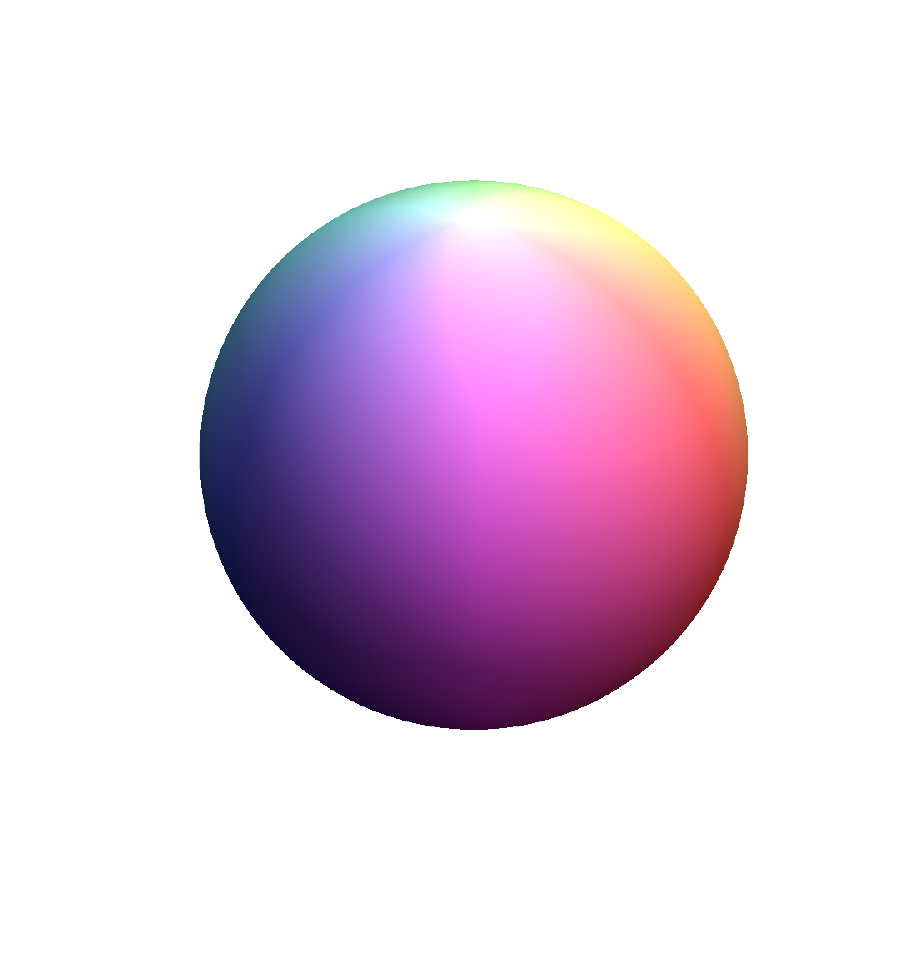}
\caption{The `hedgehog' solution.}
\label{fig:hedgehog}
\end{subfigure}
\begin{subfigure}{0.4\textwidth}
\centering
\includegraphics[height = 5cm]{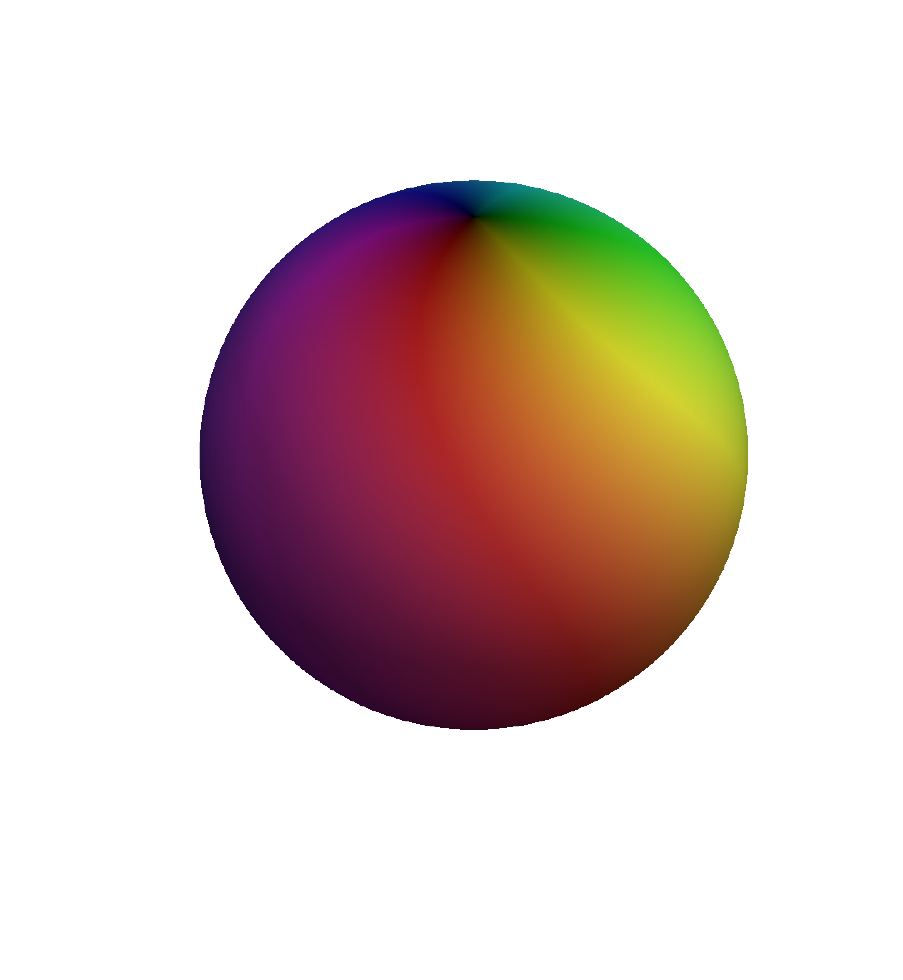}
\caption{The degree zero solution \eqref{eq:dv}.}
\label{fig:dv}
\end{subfigure}
\caption{Some Skyrmion solutions on the sphere (for \(\kappa=1\)), visualised in the Runge colour scheme. In this scheme, the argument of the complex-valued function \(v\) is represented by the hue of the colour, while the modulus is represented by the brightness (so that black corresponds to zero and white to \(\infty\)).}
\end{figure}

It would be interesting to explore more deeply the shapes of the solutions one obtains as one varies the free meromorphic function \(f\) and also as one varies the parameters of the model. 

\subsection{Skyrmions on axially symmetric films}
\label{subsec:conical}

The next step is to consider more general symmetric surfaces. Consider an axially symmetric embedding
\[
i ( r, \theta) = (r, \theta, f(r))
\]
where \((r, \theta)\) are plane polar coordinates. This becomes more complicated than the case of the sphere because explicitly finding a good complex coordinate, and computing the function \(\bar{u}_{\bar{z}}\) in \eqref{eq:boggen}, is more difficult. We compute directly, using the symmetry of the problem to simplify matters.

Using primes to indicate derivatives with respect to \(r\), the induced metric on the surface is
\[
i^* g = \left(1+f'^2\right)\dd r^2  + r^2 \dd \theta^2\text{.}
\]
The obvious complex coordinate on the surface is
\[
u = r e^{\ii\theta} \text{,}
\]
which is generally not conformal. We introduce another complex coordinate
\[
z = \tilde{r} e^{\ii\theta}
\]
where \(\tilde{r}\) is a function of \(r\) only. For this to be compatible with the complex structure on the surface induced by the embedding, we must have
\[
i^*g = \Omega^2(z,\bar{z}) \dd z \dd \bar{z}
\]
for some conformal factor \(\Omega^2\). We see that 
\[
\dd z \dd \bar{z} = \tilde{r}'^2 \dd r^2 + \tilde{r}^2 \dd \theta^2 \text{,}
\]
which is conformal to \(i^*g\) if
\[
\tilde{r}' = \frac{\sqrt{1+(f')^2}}{r} \tilde{r} 
\]
with conformal factor
\[
\frac{r^2}{\tilde{r}^2} \text{.}
\]

With this knowledge, we can write down the Bogomolny equation \eqref{eq:boggen}. On the left-hand side we have
\[
\frac{\partial}{\partial \bar{z}} v = \frac{1}{2}\frac{r}{\tilde{r}e^{-\ii\theta}}\left(\frac{1}{\sqrt{1+(f')^2}} \frac{\partial}{\partial r} + \frac{\ii}{r} \frac{\partial}{\partial \theta} \right) v \text{.}
\]
On the right, we compute 
\begin{align*}
\bar{u}_{\bar{z}} &= \frac{\partial}{\partial \bar{z}} r e^{\ii \theta} \\
	&= \frac{1}{2} \frac{r}{\tilde{r}} \frac{2}{1+|n|^2}
\end{align*}
leaving us with 
\[
\frac{1}{2}\frac{r}{\tilde{r}e^{-\ii\theta}}\left(\frac{1}{\sqrt{1+(f')^2}} \frac{\partial}{\partial r} + \frac{\ii}{r} \frac{\partial}{\partial \theta} \right) v = \ii \kappa \frac{r}{\tilde{r}} \frac{1}{1+|n|^2} \left(v-n\right)^2 \text{.}
\]
We can cancel the unknown \(\frac{r}{\tilde{r}}\) from both sides, giving
\begin{equation}
\label{eq:axbog}
\left( \frac{1-|n|^2}{1+|n|^2 }\partial_r + \frac{\ii}{r} \partial_\theta \right) v	= \ii e^{-\ii\theta} \frac{2}{1+|n|^2} \left( v - n \right)^2 \text{.}
\end{equation}

In general, the Gauss map \(n\) is not holomorphic and so \(v = n\) is not a solution in general. However, if the film is asymptotically conical, so that \(\partial_r n \to 0\) as \(r \to \infty\), then one expects there to be solutions which slowly approach the Gauss map \(v(r) \to n\) as \( r \to \infty\)  by virtue of the \(\frac{1}{r}\) suppression of the angular derivative.

The equation \eqref{eq:axbog} is not easy to solve in general. The complication, compared to the case of the sphere, is that the Gauss map is not holomorphic and so the equation does not separate. However, we can make explicit progress by looking for axially symmetric solutions for particularly simple choices of \(f\).

First, let us make the symmetric ansatz
\[
v = h(r)e^{\ii p \theta}
\]
for \(p\) an integer and \(h\) a complex-valued function (it is too much to ask that \(h\) be real, as we do not expect solutions to point radially everywhere). Substituting this into \eqref{eq:axbog}, we have
\[
\left( \frac{1-|n|^2}{1+|n|^2 }h'(r) - \frac{p}{r} h(r)\right) e^{\ii p \theta} 	= \ii e^{\ii (2p-1)\theta} \frac{2}{1+|n|^2} \left( h(r) - |n| \right)^2\text{,}
\]
where we have used the axial symmetry of the Gauss map. We see immediately that this can only be solved for \(p = 1\). In this case, we have
\begin{equation}
\label{eq:heq}
\frac{1-|n|^2}{1+|n|^2 }h'(r) - \frac{1}{r} h(r) 	= \ii \frac{2}{1+|n|^2} \left( h(r) - |n| \right)^2\text{.}
\end{equation}

The simplest example outside of the plane and the sphere is to take \(f = ar\), for a real non-zero constant \(a\). This describes a cone. Then \(|n| = \frac{a}{1+\sqrt{1+a^2}}\) does not depend on \(r\). In this case the equation \eqref{eq:heq} can be solved exactly, for any choice of \(a\). In general the solution involves (modified) Bessel functions of the first and second kind of orders determined by \(a\). 

For simplicity, we reproduce the solution here only for the value \(a = \sqrt{3}\), corresponding to an apex angle of \(\pi/3\). The solution is
\begin{align}
\label{eq:sol}
h_c(r) &= \frac{1}{2 \sqrt{2} 3^{3/4} \sqrt{-i r} \left(c I_3\left(2 \sqrt{2} \sqrt[4]{3} \sqrt{-i r}\right)-K_3\left(2 \sqrt{2} \sqrt[4]{3} \sqrt{-i r}\right)\right)} \nonumber\\
&\times \bigg(2 c I_2\left(2 \sqrt{2} \sqrt[4]{3} \sqrt{-i r}\right) + \frac{\sqrt[4]{3} \sqrt{2} c \left(\sqrt{3}-2 i r\right) I_3\left(2 \sqrt{2} \sqrt[4]{3} \sqrt{-i r}\right)}{\sqrt{-i r}} + 2 c I_4\left(2 \sqrt{2} \sqrt[4]{3} \sqrt{-i r}\right) \\
&+2 K_2\left(2 \sqrt{2} \sqrt[4]{3} \sqrt{-i r}\right) -\frac{\sqrt[4]{3} \sqrt{2}  \left(\sqrt{3}-2 i r\right) K_3\left(2 \sqrt{2} \sqrt[4]{3} \sqrt{-i r}\right)}{\sqrt{-i r}}+2 K_4\left(2 \sqrt{2} \sqrt[4]{3} \sqrt{-i r}\right)\bigg)\text{,} \nonumber
\end{align}
where \(c\) is the constant of integration, which we allow to be complex, the \(I_\alpha\) are the modified Bessel functions of the first kind and the \(K_\alpha\) are the modified Bessel functions of the second kind (note that the arguments of these functions in the solution are complex, so it would be equally reasonable to write the solution in terms of the usual Bessel functions). Introducing \(\xi = 2 \sqrt{2} \sqrt[4]{3} \sqrt{-i r}\), the solution takes the slightly more compact form:
\begin{align*}
\frac{1}{\sqrt{3}\xi \left(cI_3(\xi) - K_3(\xi)\right)} 
 \bigg(2c I_2(\xi) &+ c\left(\frac{12}{\xi} + \xi \right) I_3(\xi) + 2c I_4(\xi) \\
 &+ 2K_2(\xi) - \left(\frac{12}{\xi} + \xi \right) K_3(\xi) +2K_4(\xi)\bigg) \text{.}
\end{align*}

One can compute the limit of these solutions as \(r \to \infty\), finding that
\[
\lim_{r \to \infty} h_c(r) = \frac{1}{\sqrt{3}} = |n|\text{,}
\]
so that the magnetisation tends towards the Gauss map as \(r \to \infty\), as expected. Despite this, one should note that these solutions do not have finite energy - the solution does not tend to the normal fast enough and there is an infinite contribution to the energy from the integrated vorticity. In spite of this, we provide a short study of these solutions as they may provide a model for solutions on bumps in thin films and they exhibit an interesting confinement phenomenon. Moreover, it has been suggested in \cite{bartonMS} that it is correct to remove the integrated vorticity contribution. If one does this, the solutions we find on the cone have finite energy.



\begin{figure}
\centering
\begin{subfigure}{0.4\textwidth}
\centering
\includegraphics[height = 4cm]{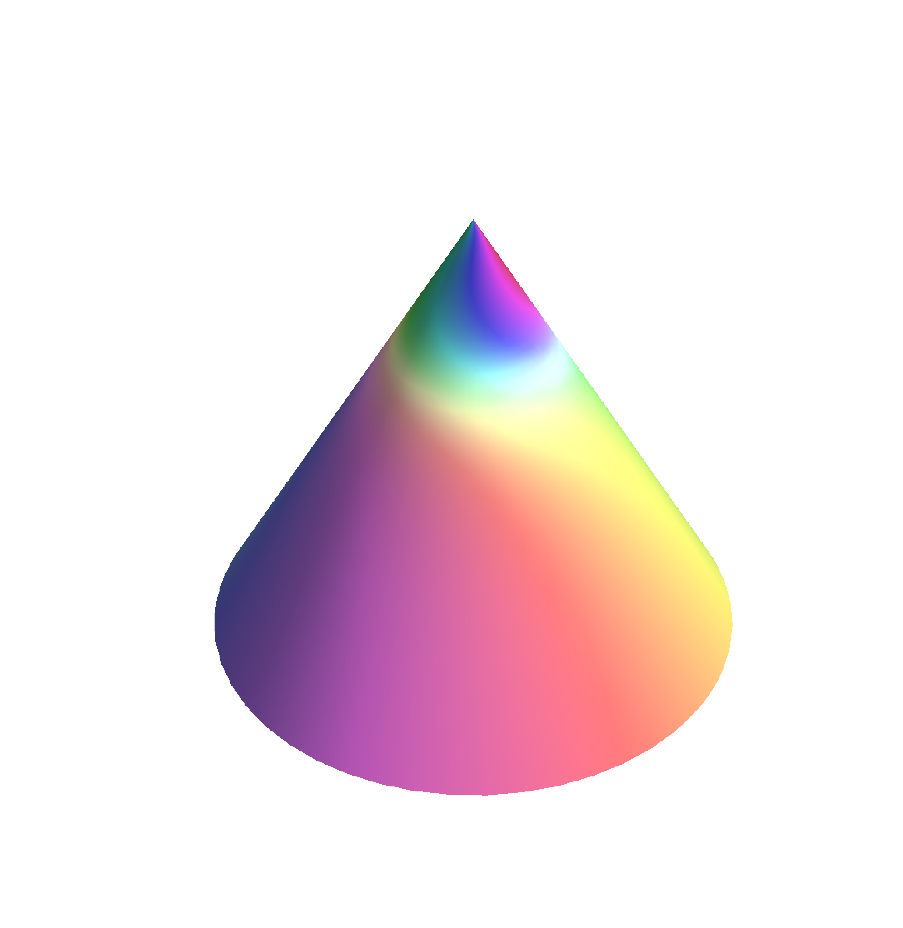}
\caption{The solution at \(c=0.1\), visualised in the Runge colour scheme.}
\label{fig:cone1}
\end{subfigure}
\begin{subfigure}{0.4\textwidth}
\centering
\includegraphics[height = 4cm]{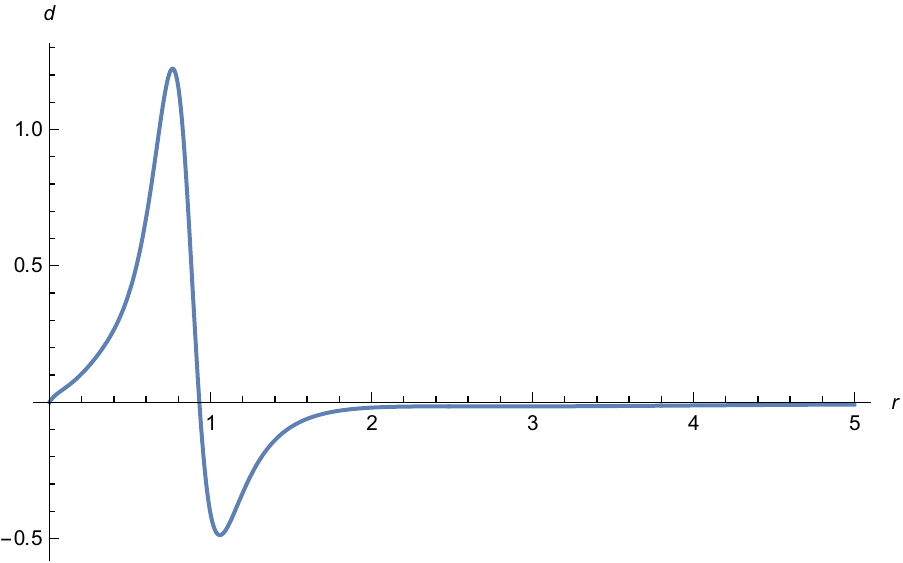}
\caption{The radial Skrymion density \(d\) of the solution at \(c=0.1\) plotted against \(r\).}
\label{fig:dens}
\end{subfigure}
\caption{Visualisations of the solution \eqref{eq:sol} at \(c=0.1\). The solution describes a ring of Skyrmion--anti-Skyrmion density around the tip of the cone.}
\label{fig:conegen}
\end{figure}

We define the radial Skyrmion density 
\[
d(r) \coloneqq   \frac{h'\bar{h} + {h}\bar{h'}}{(1+|h|^2)^2}\text{.}
\]
This is the quantity such that the `Skyrmion number' \(N\) (which may not be well-defined as an integer in this non-compact case) is the integral \(\int  d(r) \, \dd r \). In \autoref{fig:dens}, we plot \(d\) against \(r\) (in general, \(r\) is not the right coordinate to use, but for the cone it differs from the radial distance along the cone just by a scaling), and see that the solution describes a band of Skyrmion--anti-Skyrmion density around the tip of the cone.

\begin{figure}
 \centering
\begin{subfigure}{0.45\textwidth}
\includegraphics[width=\linewidth, height=5cm]{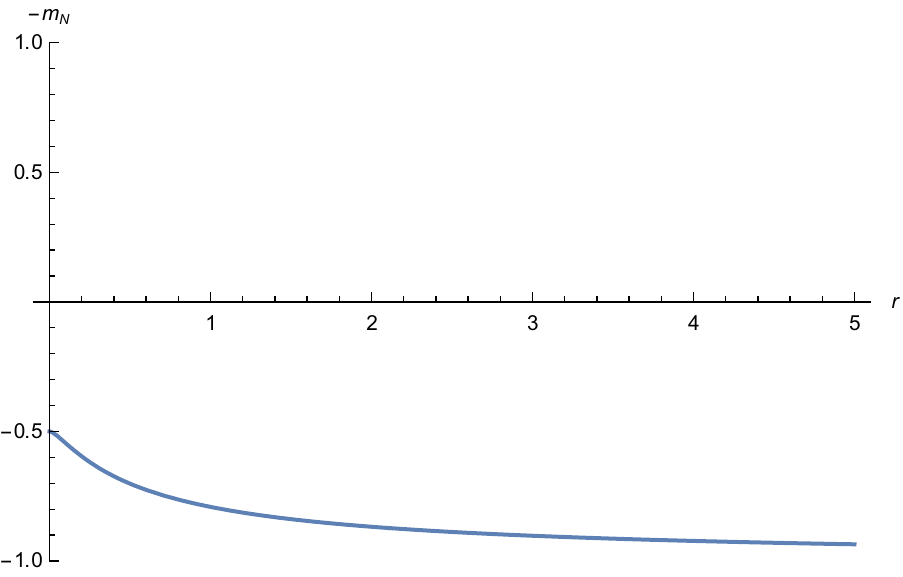} 
\caption{}
\label{fig:sol0}
\end{subfigure}
\begin{subfigure}{0.45\textwidth}
\includegraphics[width=\linewidth, height=5cm]{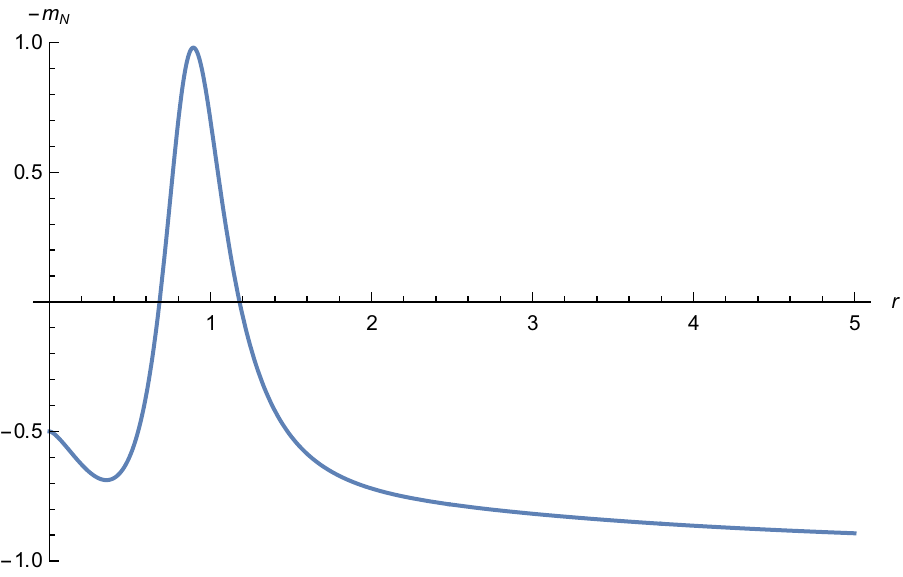}
\caption{}
\label{fig:sol0.5}
\end{subfigure}

\begin{subfigure}{0.45\textwidth}
\includegraphics[width=\linewidth, height=5cm]{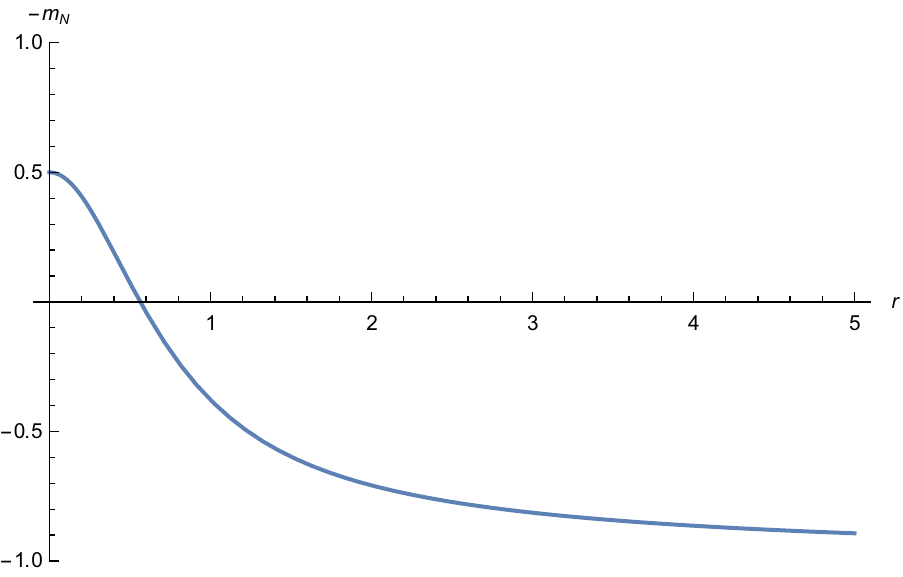} 
\caption{}
\label{fig:solinf}
\end{subfigure}
\begin{subfigure}{0.45\textwidth}
\includegraphics[width=\linewidth, height=5cm]{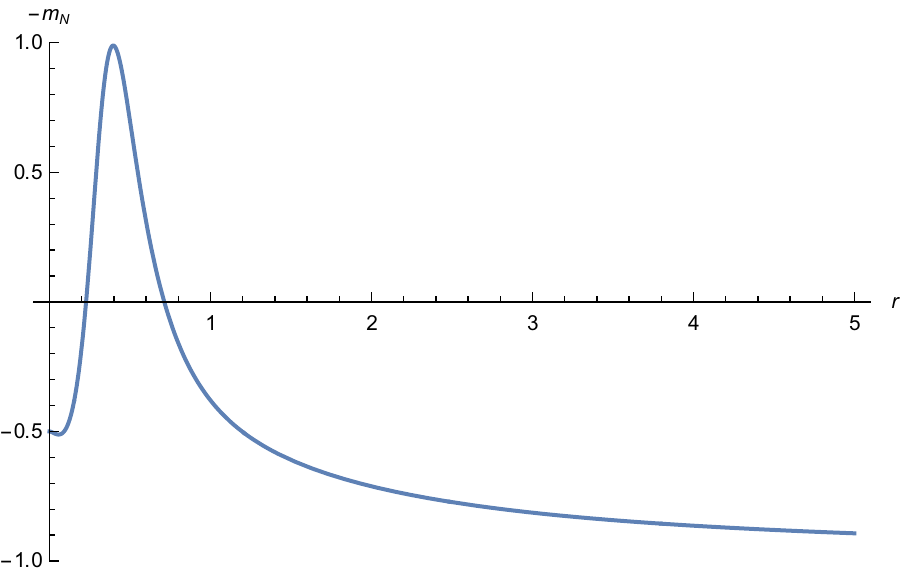} 
\caption{}
\label{fig:soli}
\end{subfigure}

\caption{The negative normal component of the magnetisation field, \(-m_N\), against the radial distance \(r\) for axially symmetric solutions of the form \eqref{eq:sol} on the cone of slope \(a=\sqrt{3}\) at various values of the parameter \(c\): (a) \(c=0\), (b) \(c=0.1\), (c) \(c = \infty\), and (d) \(c = -\ii\).}
\label{fig:norm}
\end{figure}

To understand how the solution varies as one varies the modulus \(c\),  we plot in \autoref{fig:norm} the negative of the normal component of the magnetisation vector field, \(-m_N\), against \(r\) for several values of \(c\). 

Among the values we have plotted are the two special values \(c=0\) and \(c=\infty\). The case of \(c=0\) describes the natural `ground state', where the system remains as close as possible to the normal vector field (note that at \(r=0\), the `normal' would point straight along the axis of the cone, which corresponds -- for the cone angle we have chosen -- to \(m_N = 0.5\)). We have also plotted the case of \(c = 0.1\), which can be compared to \autoref{fig:conegen}.

On the other hand, the solution for \(c=\infty\) describes a pointlike Skyrmion at the tip of the cone (again, a value of \(m_N = - 0.5\) at \(r=0\) implies that the field points along the axis of the cone in the `anti-normal' direction there). This is the special case in which the radius of the ring of Skyrmion density goes to zero.

It was demonstrated in \cite{kravchukSCD} that Skyrmions can be pinned in place by bumps in a thin film. It would be desirable to take our analysis further, by studying the theory on a smooth, asymptotically Euclidean `bump'. Unfortunately, reasonable choices of bump lead to Bogomolny equations that seem to be very difficult to solve. One could instead try to make progress numerically, although we do not do so here. 

\subsection{Skyrmions on cylindrical films}

Perhaps more interesting than the case of conical films is that of cylindrical films. The parameterisation that we used above does not naturally allow for cylindrical films, so we treat them separately here. Skyrmions on cylinders may be of interest in the study of nanowires.

A symmetric embedding of a cylinder in \(\mathbb{R}^3\) around the \(x_3\)-axis takes the form
\[
i(t, \theta) = (R\cos \theta, R\sin \theta, t)
\]
where \(t \in \mathbb{R}\), \(\theta \in [0, 2\pi)\) and \(R\) is a positive constant. The induced metric on the film is
\[
g_\Sigma = \dd t^2 + R^2\dd \theta^2\text{.}
\]
The local complex coordinate \(z = t + \ii R\theta\) is then conformal.

A short computation, along the lines of those we have already carried out, reveals that the Bogomolny equation \eqref{eq:bogv} is
\[
\frac{\partial v}{\partial t} + \frac{\ii}{R} \frac{\partial v}{\partial \theta} = \ii e^{-\ii \theta} \left(v - e^{\ii \theta}\right)^2\text{.}
\]
Note that the Gauss map of the cylinder is \(n = e^{\ii \theta}\), so this equation takes a similar form to those we have seen previously.

Again making the axial ansatz \(v = h(t) e^{\ii \theta}\), we find the separable equation
\[
h'(t) = \frac{1}{R}h(t) + \ii(h(t)-1)^2
\]
which can be solved to give
\begin{equation}
\label{eq:cylsol}
h_c(t) =\frac{i \sqrt{1-4 i R}}{2R} \tanh \left(\frac{1}{2} \left(\sqrt{1-4 i R}\left(c+t/R\right) \right)\right) + \frac{\ii}{2R} + 1
\end{equation}
which depends on the complex parameter \(c\). This describes a `kink', mediating between the asymptotic constant solutions
\begin{equation}
\label{eq:cylasymp}
h_\pm = \frac{1}{2R}\left( \ii + 2R \pm {i \sqrt{1-4 i R}}\right)
\end{equation}
which solve the \(t\)-independent Bogomolny equation
\[
\frac{1}{R}h + \ii (h-1)^2 = 0 \text{.} 
\]
Notice that these asymptotic solutions do not describe the Gauss map for finite \(R\). Because the `asymptotic boundary' of the cylinder at \(t \to \pm\infty\) has finite length \(2\pi R\), this does not necessarily cause issues when evaluating the energy contribution of the integrated vorticity. The \(t\)-independent solutions \eqref{eq:cylasymp} are true vacua: they have vanishing Skyrmion density and the integrated vorticity vanishes, as the contribution from each end of the cylinder cancels, and so have zero energy. This occurs in spite of the fact that they have positive potential energy and gradient energy, because the DM energy contributes negatively. 

\begin{figure}[h!]
 \centering
\begin{subfigure}{0.45\textwidth}
\includegraphics[width=\linewidth, height=5cm]{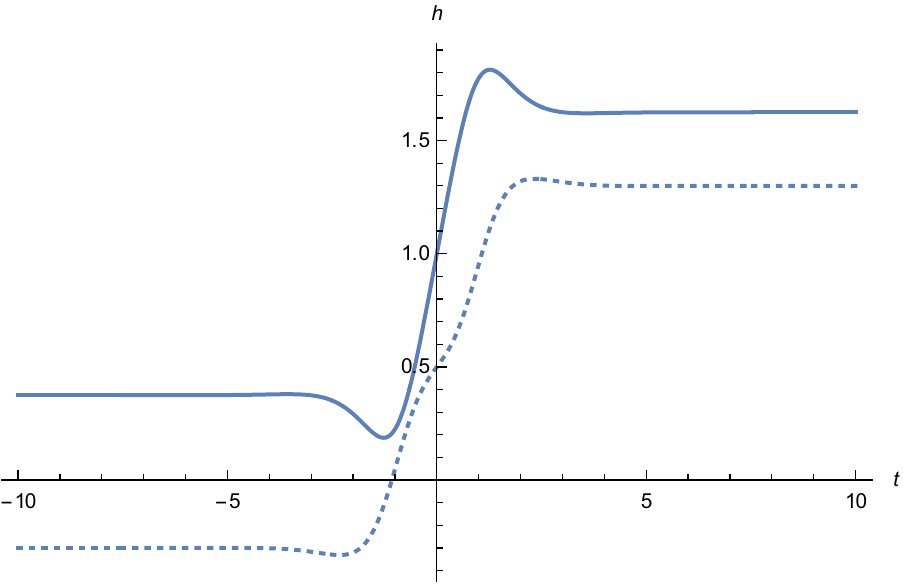} 
\caption{}
\label{fig:cylsol0}
\end{subfigure}
\begin{subfigure}{0.45\textwidth}
\includegraphics[width=\linewidth, height=5cm]{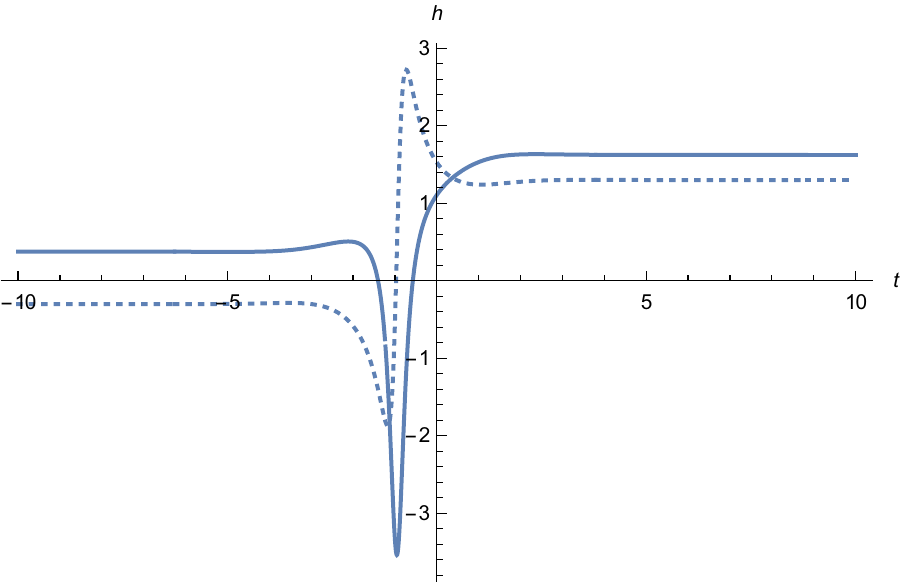}
\caption{}
\label{fig:cylsoli}
\end{subfigure}

\begin{subfigure}{0.45\textwidth}
\includegraphics[width=\linewidth, height=5cm]{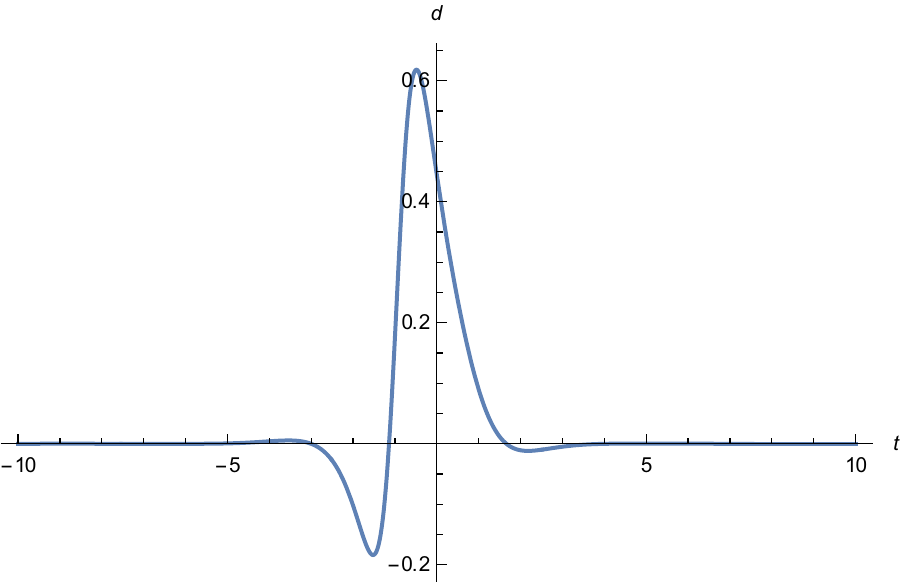} 
\caption{}
\label{fig:cyldens0}
\end{subfigure}
\begin{subfigure}{0.45\textwidth}
\includegraphics[width=\linewidth, height=5cm]{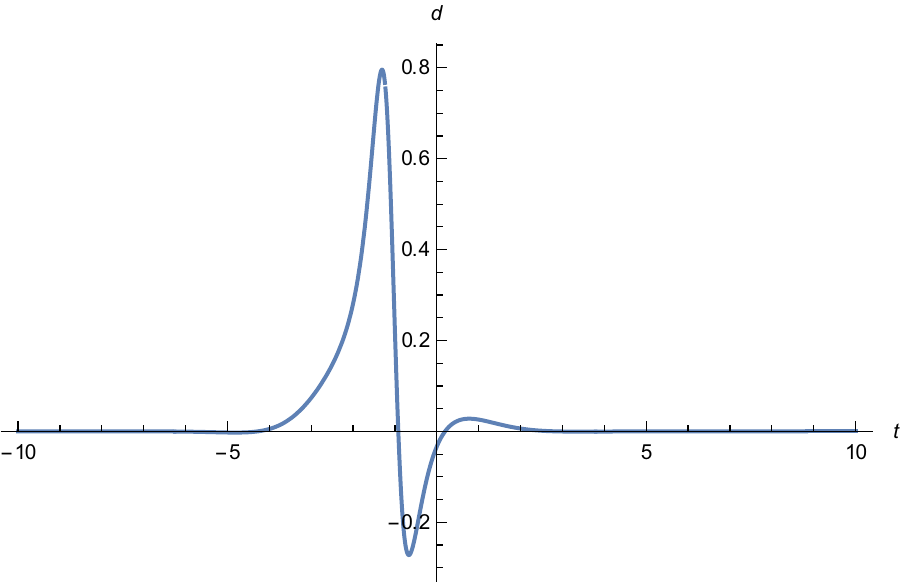} 
\caption{}
\label{fig:cyldensi}
\end{subfigure}

\caption{The figures (a) and (b) are plots of the real (solid line) and imaginary parts (dashed line) of \(h_c(t)\) at \(c=0\) and \(c=i\) respectively for \(R=1\). Figures (c) and (d) are the corresponding plots of Skyrmion density \(d\) against \(t\) for the solutions \(h_0\) and \(h_i\) respectively. Notice that changing \(c\) moves the centre of the kink and changes its shape.}
\label{fig:cylsols}
\end{figure}

The complex modulus \(c\) controls the position and shape of the kink. In \autoref{fig:cylsols}, we plot some solutions and their Skyrmion density profiles.

By evaluating the integrated Skyrmion density and integrated vorticity, one can compute the energy of a Skyrmion kink solution. One finds that it is finite and does not depend on \(c\). However, it depends on the radius \(R\) of the cylinder. 

Indeed, by examining the solution \eqref{eq:cylsol}, one can see that for large \(R\), the integrated Skyrmion density goes like \(1/\sqrt{R}\) (note that the asymptotic solution as \(R\to \infty\) is the Gauss map which has no Skyrmion density), while the integrated vorticity goes like \(\sqrt{R}\) (it behaves like \(R(h_+ - h_-)\)). One can further compute the constants, and finds that the integrated Skyrmion density has leading term
\[
4\pi \frac{2}{\sqrt{R}}
\]
as \(R\to \infty\), while the integrated vorticity has leading term
\[
4\pi \frac{\sqrt{R}}{2}
\]
as \(R\to \infty\). (Recall that we are working in units where the Skyrmion size \(\frac{1}{\kappa}\) has been set to one.)

\section{The moduli space of Skyrmions and its resolution}
\label{sec:moduli}

\subsection{The moduli space of Skyrmions}

One of the crucial facts about the critically coupled model is that it admits a \emph{moduli space} of degenerate energy minimising solutions within each topological class. These moduli include the positions of the basic Skyrmions (which exert no net force on each other at critical coupling) as well as certain internal `orientational' moduli. 

The moduli space of solutions to the Bogomolny equation \eqref{eq:bog} is the space of sections of a (topologically trivial) projective bundle with holomorphic structure determined by \(A^{0,1}\). Let us restrict to the case that \(\Sigma\) is compact without boundary. An index theory argument shows that the component of the moduli space corresponding to sections of degree \(N\) has expected complex dimension
\[
2N + 1 - g
\]
where \(g\) is the genus of \(\Sigma\). This is the same as the expected dimension of the space of meromorphic functions on  \(\Sigma\).

The moduli space of sections of a projective bundle contains singular points, at which the section degenerates to one of lower degree. From the point of view of Skyrmions, these singularities are points at which a Skyrmion shrinks to zero size and disappears. The aim of this section is to resolve these singularities.

\subsection{Semi-local vortices in a background}

To resolve the singularities of the moduli space we may use the idea, well-known in the literature of high energy physics, of realising a nonlinear sigma model as a certain strong coupling limit of a gauge theory with linear Higgs fields. The Skyrmions of the nonlinear theory are then the limits of vortices in the gauge theory. The vortices of the gauge theory have a well-defined size and a non-singular moduli space. We may, at least formally, do computations on the vortex moduli space and then take the strong coupling limit to find information about the Skyrmion theory. 

From a two-dimensional perspective, the theory we consider is an Abelian gauge theory with a pair of complex scalar Higgs fields coupled to a background gauge potential for the \(SU(2)\) flavour symmetry, which rotates the scalar fields among themselves. However, just as before, our theory has a three-dimensional origin - really the two scalar fields arise as the restriction of a spinor field in three dimensions and the background gauge field comes from a spin connection with torsion.  The coupling to this connection and the relation to the theory of magnetic Skyrmions are the only potential novelties in the following discussion.

Let \(M\) be a three-dimensional Riemannian manifold. To define the vortex theory we will replace the magnetisation vector field \(m\) with a spinor on \(M\).  To do this, we assume that \(M\) is oriented and let \(S \to M\) be a spinor bundle. This is a Hermitian complex vector bundle of rank 2, with vanishing Chern class. 

We again let \(i: \Sigma \hookrightarrow M\) be an embedded thin film. We introduce a Hermitian complex line bundle \(L \to \Sigma\) and define \(E\coloneqq i^*S \otimes L\). The fields of our gauge theory are then a section \(\Psi \in \Gamma(E)\) and a \(U(1)\) connection \(a\) on \(L\). The  background \(SO(3)\) connection \(A\) can be lifted to an \(SU(2)\) connection on \(i^*S\) which we also call \(A\). We may then form the tensor product \(U(2)\) connection 
\[
 A\otimes 1 + \mathbbm{1} \otimes a
\]
on \(E\), which we write as \(A+a\).

We introduce the energy functional
\[
E_{A,e^2} = \frac{1}{2}\int_\Sigma * \left( |\dd_{A+a} \Psi |^2 + \frac{1}{e^2}|F(a)|^2 + e^2(v^2-|\Psi|^2)^2 - * 2\Psi^\dagger F(A) \Psi \right)\text{.}
\]
This functional depends on two positive real parameters: the `saturation magnetisation' \(|v|\), which may as well be set to \(1\), and the gauge coupling constant \(e^2\) (which may be thought of as a choice of inner product on the Lie algebra of \(U(1)\)). 

A well-known rearrangement (see \ref{sec:nlsm}) shows that, up to boundary terms,
\[
E_{A,e^2} = \int_\Sigma \left( | \db_{A+a} \Psi |^2 + \frac{1}{e^2} | F(a) - \ii e^2 (v^2-|\Psi|^2) |^2 \right) +  \int_\Sigma  \ii v^2 F(a) \text{.}
\]
Thus, the energy is minimised within a topological class by solutions to the \emph{vortex equations}
\begin{align}
\db_{A+a} \Psi &= 0 \label{eq:vor1}\\
*F(a) &= \ii e^2(v^2 - |\Psi|^2) \label{eq:vor2}\text{.}
\end{align}
Exact solutions to equations of this type are usually very hard to come by. However, the moduli space of solutions to the analogue of these equations with \(A=0\), in the case that \(E\) is replaced with an arbitrary sum of line bundles (the case of  \emph{Abelian semi-local vortices}), is well-studied, as is the case in which \(A\) is dynamical and \(E\) is general (the case of \emph{non-Abelian vortices}).

The moduli space is a K\"ahler manifold. A simple index calculation, using the Riemann--Roch theorem, tells us that the expected complex dimension of the moduli space is 
\[
c_1(E)[\Sigma] + 2 - 2g + (g - 1) = 2N_L + 1 - g
\]
where \(c_1(E)\) is the first Chern class of \(E\), \(N_L\) is the degree of \(L\), and \(g\) is the genus of \(\Sigma\).

If one takes the strong coupling limit \(e^2 \to \infty\), any finite energy configuration must obey
\[
v^2 = |\Psi|^2 \text{,}
\]
which requires that \(\Psi\) takes values in a three-sphere.
In this limit, the Abelian gauge field \(a\) decouples. When we take the quotient by the \(U(1)\) gauge group we are left with a nonlinear sigma model into 
\[
S^2 = S^3/U(1) \text{,}
\]
the \emph{Higgs branch} of vacua of the gauge theory. The theory that results is precisely the critical magnetic Skyrmion theory (indeed, this point of view on the Skyrmion theory is exactly that of \autoref{subsec:proj}). 

The vortex theory at finite \(e^2\) is such that the moduli space resolves that of magnetic Skyrmions. In particular, the vortex moduli space is non-singular and, if \(\Sigma\) is closed, closed. Vortices have a size, they take up an area proportional to 
\[
\frac{1}{v^2e^2} \text{,}
\]
as can be seen by integrating \eqref{eq:vor2} over the surface \(\Sigma\). In the limit \(e^2 \to \infty\), this goes to zero. 

We can construct the vortex moduli space, as a manifold, by using instead the `dissolving vortex' limit (see \cite[§7]{wehrheimVI} and \cite{baptistaDV}). One may think of this as the weak coupling limit, where \(e^2 \to 0\), although one must be careful to enforce the topological constraints of the Bogomolny equations. To do this, we first fix the vortex number \(N\) and introduce \(\tilde{v}^2 = {v}^2 -  \frac{2\pi N}{e^2\text{vol}(\Sigma)}\) so as to rewrite the equations \eqref{eq:vor1}, \eqref{eq:vor2} as
\begin{align*}
\db_{A+a} \Psi &= 0 \\
*F(a) - \frac{2\pi \ii N}{\text{vol}(\Sigma)}  &= \ii e^2(\tilde{v}^2 - |\Psi|^2)\text{.}
\end{align*}
Integrating the second of these equations tells us that, for any \(e^2\neq 0\),
\[
\int_\Sigma * |\Psi|^2  = \tilde{v}^2\text{vol}(\Sigma)\text{.}
\]

It turns out that the system remains well-behaved as \(e^2 \to 0\) \cite{wehrheimVI}. In this limit, the vortex equations become the \emph{dissolving vortex equations}:
\begin{align}
\db_{A+a} \Psi &= 0 \label{eq:dv1}\\
*\frac{1}{2\pi\ii} F(a) &= \frac{N}{\text{vol}(\Sigma)} \label{eq:dv2} \\
\frac{1}{\text{vol}(\Sigma)}\int_\Sigma * |\Psi|^2  &= \tilde{v}^2\text{.} \label{eq:dv3}
\end{align}
So solutions consist of projectively flat line bundles \((L,a)\) equipped with a normalised holomorphic section \(\Psi\) of the holomorphic bundle \((S\otimes L, (A \otimes a)^{0,1} )\). We expect that, as a complex manifold, the moduli space of dissolving vortices is isomorphic to the moduli space of vortices at positive \(e^2\). Of course, these spaces come with a natural Riemannian structure, which does vary with \(e^2\).

Let us construct the moduli space of dissolving vortices. First, there is a one-to-one correspondence between projectively flat line bundles on a Riemann surface and holomorphic line bundles on the surface, modulo equivalence. Letting \(\text{Pic}^N(\Sigma)\) be the Picard group, the moduli space of holomorphic line bundles on \(\Sigma\) of degree \(N\), which is a torus of complex dimension \(g\), we form the universal holomorphic bundle
\[
\mc{V}_N \to \Sigma \times \text{Pic}^N(\Sigma)
\]
which has as fibre over \(\Sigma \times \{\mc{L}\}\) the bundle \(\mc{S} \otimes \mc{L}\), where \(\mc{S}\) is the pulled back spinor bundle \(i^*S\) with holomorphic structure defined by \(A^{0,1}\). We fix the bundle \(\mc{V}_N\) by choosing a point \(x \in \Sigma\) and asking that \({\mc{V}_N}|_{\{x\}\times \text{Pic}^N(\Sigma)}\) be trivial.

Regarding \(\mc{V}_N\) as a locally free sheaf, we may push it forward along the projection map \(p: \Sigma \times \text{Pic}^N(\Sigma) \to \text{Pic}^N(\Sigma)\). If \(N\) is sufficiently large, Serre vanishing tells us that the resulting sheaf is locally free of constant rank
\[
2N + 2 - 2g 
\]
(explicit bounds on \(N\) may be accessible if one computes the holomorphic type of \(\mc{S}\)). In this case, the pushforward defines a vector bundle over \(\text{Pic}^N(\Sigma)\) with fibre over \(\mc{L}\) the space of holomorphic sections of \(\mc{S}\otimes \mc{L}\). 

To impose the normalisation condition \eqref{eq:dv3} and to take the quotient of the action of gauge transformations on \(\Psi\), we should take the projectivisation of this vector bundle. The moduli space is then the total space of the corresponding projective bundle. Provided \(N\) is sufficiently large, this is a compact complex manifold of dimension 
\[
2N +1 - g
\]
as we expected.

In the case that \(g = 0\), this construction is particularly simple. In that case, the Picard group is a point and the moduli space is simply
\[
\mathbb{C}P^{2N + 1}\text{.}
\]
Notice that this resolves the moduli space of rational maps, which is
\[
\mathbb{C}P^{2N+1} - \Delta
\]
where \(\Delta\) is the resultant hypersurface, consisting of rational maps which degenerate to a lower degree map.

Note that, at least for \(N\) sufficiently large relative to \(g\), the background connection \(A\) does not affect the resolved moduli space, at least as a manifold. 

\section{Conclusions}

Modelling chiral interactions in magnetic materials by coupling the magnetisation field to a metric connection with torsion is mathematically natural and leads straightforwardly to standard DM-type interaction terms. Using this idea to understand Skyrmions on magnetic thin films leads us to the viewpoint of \cite{bartonMS}, thinking of the chiral energy functional as that of a sigma model coupled to the background torsional gauge field. In this note, we have taken this idea further by picking a physically natural  torsional connection in three dimensions and restricting it to general curved thin films. 

The further great insight of \cite{bartonMS,schroersGSM} was that, by picking a special potential function, one obtains a BPS-type theory. We have shown that solutions to the corresponding Bogomolny equation always exist on general compact thin films. Moreover, we have seen that, in certain symmetric cases, the Bogomolny equation can be solved exactly. It would be interesting to explore these exact solutions in greater depth, and to expand the class of films on which exact solutions are known. 

While magnetic Skyrmions are usually sufficiently large to be treated classically, there has been interest in very small Skyrmions where quantum effects become important (see \cite{roldanQF,psaroudakiQD}, for example). In constructing the resolved moduli space of Skyrmions, we have taken the first step towards understanding the low energy quantum dynamics (and thermodynamics) of BPS Skyrmions on curved films. In general, classical Skyrmion dynamics is governed by the Landau--Lifshitz--Gilbert equation, which combines Hamiltonian and dissipative dynamics. At sufficiently low energies, the dissipative contribution disappears and one is left with (trivial) Hamiltonian dynamics on the moduli space. Hence, the geometric quantisation of the moduli space captures the low energy quantum dynamics of these BPS Skyrmions. Formally, one can do exact computations on the vortex moduli space of \autoref{sec:moduli} at finite gauge coupling \(e^2\)  and then take the limit \(e^2 \to \infty\) to obtain information about Skyrmions.

\section*{Acknowledgements}

I am very grateful to Nick Manton for numerous useful discussions, comments and suggestions. I am also thankful to Calum Ross, for sparking my interest in the subject and for useful comments, and to Bernd Schroers, Bruno Barton-Singer, Giovanni Di Fratta, Valeriy Slastikov, and David Tong for helpful discussions. This work has been supported by an EPSRC studentship. It has also been partially supported by STFC consolidated grant ST/P000681/1.

\appendix

\section{Nonlinear sigma models in a background gauge field}
\label{sec:nlsm}

We recall some aspects of the theory of sigma models in a background gauge field. The general set up is as follows. Let \(G\) be a compact Lie group with Lie algebra \(\mathfrak{g}\) with a given Killing form. Let \((X, \omega_X)\) be a symplectic manifold with a compatible almost complex structure \(J_X\). Let \(X\) carry a Hamiltonian \(G\)-action. Physically, \(X\) will be the target space and \(G\) will be the gauge group. 

Associated to the Hamiltonian action of \(G\) on \(X\) is an equivariant moment map 
\[
\mu : X \to \mathfrak{g}^\vee 
\]
with the (defining) property that the equivariant 2-form
\[
\omega_X - \mu \in \Omega^2_G(X)
\]
is equivariantly closed \cite{atiyahMM}, which is to say that
\[
\dd \mu(\xi) = \omega(v_\xi)
\]
for all \(\xi \in \mathfrak{g}\), where \(v_\xi\) is the fundamental vector field on \(X\) associated to \(\xi\). Hence, \(\omega_X - \mu\) determines a class \([\omega_X - \mu]\) in the equivariant cohomology \(H^2_G(X)\).

Now, let \(\Sigma\) be a compact Riemannian 2-manifold, playing the role of physical space. We would like to set up a background gauge theory, and so we take a principal \(G\)-bundle \(P \to \Sigma\). Our nonlinear field is a gauged map into the target space \(X\) (i.e. an \(X\)-valued field, charged under \(G\)), which is to say that it is a section of the associated bundle
\[
\ul{X} \coloneqq X \times_G P \to \Sigma \text{.}
\]
A section of this bundle is equivalently a \(G\)-equivariant map \(P \to X\).

Now, we should introduce a background gauge field, which is a connection \(A\) on \(P\). This allows us to define the covariant derivative \(\dd_A \phi \in \Omega^2_V(\ul{X})
\) of a section \(\phi \in \Gamma(\ul{X})\). Here, by \(\Omega^2_V(\ul{X})\) we mean the space of vertical 2-forms. We also define the antiholomorphic derivative
\[
\db_A \phi = \frac{1}{2}\left(\dd_A \phi + J_X \circ \dd_A\phi \circ j_\Sigma \right)\text{.}
\]

The energy functional we consider is the usual one:
\[
E[A][\phi] = \frac{1}{2} \int_\Sigma \left(|\dd_A \phi|^2 + V(\phi, A)) \right)\text{vol}_\Sigma\text{.}
\]
The critically coupled model arises for a special potential, namely,
\begin{equation}
V_\text{crit}(\phi, A) =  -2 \mu(\phi) ( * F(A)) \label{eq:crit}
\end{equation}
where \(*\) is the Hodge star on \(\Sigma\). The equivariance of the moment map implies that this potential is gauge invariant.

The reason that  this is called `critically coupled' is that, for this choice of potential, the energy admits a Bogomolny rearrangement. Indeed, it was shown in \cite{cieliebakSV} that
\begin{equation}
\label{eq:generalbog}
\frac{1}{2}\int_\Sigma * \left( |F(A)|^2 + | \dd_A \phi |^2 + | \mu (\phi ) |^2 \right) = \int_\Sigma *\left( |\db_A \phi|^2 + \frac{1}{2}| *F(A) + \mu^\sharp(\phi) |^2 \right) + [\omega - \mu] ( [\phi])
\end{equation}
where the second term on the right-hand side is the natural pairing of \([\omega_X - \mu] \in H^2_G(X)\) with \([\phi] \in H_2^G(X)\) and \(\mu^\sharp(\phi)\) is the moment map with its index raised using the Killing form, so that it takes values in \(\mathfrak{g}\). This rearrangement is of interest in the study of (symplectic) vortices.

Rearranging \eqref{eq:generalbog}, we see that
\begin{align*}
\int_\Sigma * |\db_A \phi |^2 + [\omega - \mu] ( [\phi]) &= \frac{1}{2} \int_\Sigma * \left( | \dd_A \phi |^2 + |F(A)|^2 + |\mu(\phi)|^2 - |*F(A) + \mu^\sharp (\phi)|^2 \right) \\
	&= \frac{1}{2} \int_\Sigma * \left( | \dd_A \phi |^2 - 2 \mu(\phi) ( *F(A)) \right)\text{.}
\end{align*}
The right-hand side is our energy functional.
The energy of a configuration is therefore bounded below within a topological sector by the topological energy
\[
[\omega - \mu] ( [\phi])
\]
and the bound is saturated by configurations \(\phi\) obeying
\[
\db_A \phi = 0 \text{.}
\]

For the model studied in this paper, we take \(X\) to be the sphere \(S^2\) with its standard symplectic form \(\omega_{S^2}\). We take \(G=SO(3)\), which acts on \(S^2\) by rotations in the usual way preserving \(\omega_{S^2}\). We now write \(m\) for the field \(\phi\), as we now interpret it as the magnetisation field. It can be shown that the moment map for this action 
\[
\mu : S^2 \to \mathfrak{so}(3)
\]
is the inclusion of the unit sphere. Hence, thinking of \(m= (m^1,m^2,m^3)\) as a unit vector in \(\mathbb{R}^3 \cong \mathfrak{so}(3)\), we have
\[
\mu (m ) (\xi ) = m^B\xi_B
\]
for \(\xi \in \mathfrak{so}(3)\) and where \(B=1,2,3\) labels coordinates on \(\mathfrak{so}(3)\). Substituting this into the critical potential \eqref{eq:crit} leads to the potential term \eqref{eq:cpot}.

\bibliographystyle{elsarticle-num}
\bibliography{bibliography.bib}

\begin{thebibliography}{10}
\expandafter\ifx\csname url\endcsname\relax
  \def\url#1{\texttt{#1}}\fi
\expandafter\ifx\csname urlprefix\endcsname\relax\def\urlprefix{URL }\fi
\expandafter\ifx\csname href\endcsname\relax
  \def\href#1#2{#2} \def\path#1{#1}\fi

\bibitem{polyakovMS}
A.~M. Polyakov, A.~A. Belavin, {Metastable States of Two-Dimensional Isotropic
  Ferromagnets}, JETP Lett. 22 (1975) 245--248.

\bibitem{bogdanovMVS}
A.~Bogdanov, A.~Hubert, Thermodynamically stable magnetic vortex states in
  magnetic crystals, J. Magn. Magn. Mater. 138~(3) (1994) 255--269.
\newblock \href {https://doi.org/10.1016/0304-8853(94)90046-9}
  {\path{doi:10.1016/0304-8853(94)90046-9}}.

\bibitem{muhlbauerMS}
S.~M{\"u}hlbauer, B.~Binz, F.~Jonietz, C.~Pfleiderer, A.~Rosch, A.~Neubauer,
  R.~Georgii, P.~B{\"o}ni, Skyrmion lattice in a chiral magnet, Science
  323~(5916) (2009) 915--919.
\newblock \href {https://doi.org/10.1126/science.1166767}
  {\path{doi:10.1126/science.1166767}}.

\bibitem{yuSC}
X.~Yu, Y.~Onose, N.~Kanazawa, J.~Park, J.~Han, Y.~Matsui, N.~Nagaosa,
  Y.~Tokura, Real-space observation of a two-dimensional {S}kyrmion crystal,
  Nature 465~(7300) (2010) 901.
\newblock \href {https://doi.org/10.1038/nature09124}
  {\path{doi:10.1038/nature09124}}.

\bibitem{bartonMS}
{B. Barton-Singer}, {C. Ross}, {B. J. Schroers}, {Magnetic Skyrmions at
  Critical Coupling}, Commun. Math. Phys. (online). (2020).
\newblock \href {https://doi.org/10.1007/s00220-019-03676-1}
  {\path{doi:10.1007/s00220-019-03676-1}}.

\bibitem{schroersGSM}
B.~J. Schroers, {Gauged Sigma Models and Magnetic Skyrmions}, SciPost Phys. 7
  (2019) 30.
\newblock \href {https://doi.org/10.21468/SciPostPhys.7.3.030}
  {\path{doi:10.21468/SciPostPhys.7.3.030}}.

\bibitem{hongoICM}
M.~Hongo, T.~Fujimori, T.~Misumi, M.~Nitta, N.~Sakai, Instantons in chiral
  magnets, Phys. Rev. B 101 (2020) 104417.
\newblock \href {https://doi.org/10.1103/PhysRevB.101.104417}
  {\path{doi:10.1103/PhysRevB.101.104417}}.

\bibitem{gaidideiCE}
Y.~Gaididei, V.~P. Kravchuk, D.~D. Sheka, Curvature effects in thin magnetic
  shells, Phys. Rev. Lett. 112 (2014) 257203.
\newblock \href {https://doi.org/10.1103/PhysRevLett.112.257203}
  {\path{doi:10.1103/PhysRevLett.112.257203}}.

\bibitem{streubelMCG}
R.~Streubel, P.~Fischer, F.~Kronast, V.~P. Kravchuk, D.~D. Sheka, Y.~Gaididei,
  O.~G. Schmidt, D.~Makarov, Magnetism in curved geometries, J. Phys. D 49~(36)
  (2016) 363001.
\newblock \href {https://doi.org/10.1088/0022-3727/49/36/363001}
  {\path{doi:10.1088/0022-3727/49/36/363001}}.

\bibitem{kravchukSCD}
V.~P. Kravchuk, D.~D. Sheka, A.~K\'akay, O.~M. Volkov, U.~K. R\"o\ss{}ler,
  J.~van~den Brink, D.~Makarov, Y.~Gaididei, Multiplet of {S}kyrmion states on
  a curvilinear defect: Reconfigurable {S}kyrmion lattices, Phys. Rev. Lett.
  120 (2018) 067201.
\newblock \href {https://doi.org/10.1103/PhysRevLett.120.067201}
  {\path{doi:10.1103/PhysRevLett.120.067201}}.

\bibitem{hsuSMC}
L.~Hsu, R.~Kusner, J.~Sullivan, Minimizing the squared mean curvature integral
  for surfaces in space forms, Exp. Math. 1~(3) (1992) 191--207.
\newblock \href {https://doi.org/10.1080/10586458.1992.10504258}
  {\path{doi:10.1080/10586458.1992.10504258}}.

\bibitem{mildeUSL}
P.~Milde, D.~K{\"o}hler, J.~Seidel, L.~M. Eng, A.~Bauer, A.~Chacon,
  J.~Kindervater, S.~M{\"u}hlbauer, C.~Pfleiderer, S.~Buhrandt, C.~Sch{\"u}tte,
  A.~Rosch, Unwinding of a {S}kyrmion lattice by magnetic monopoles, Science
  340~(6136) (2013) 1076--1080.
\newblock \href {https://doi.org/10.1126/science.1234657}
  {\path{doi:10.1126/science.1234657}}.

\bibitem{kanazawaCP}
{N. Kanazawa}, {Y. Nii}, {X. -X. Zhang}, {A. S. Mishchenko}, {G. De Filippis},
  {F. Kagawa}, {Y. Iwasa}, {N. Nagaosa}, {Y. Tokura}, {Critical phenomena of
  emergent magnetic monopoles in a chiral magnet}, Nat. Comm. 7 (2016) 11622.
\newblock \href {https://doi.org/https://doi.org/10.1038/ncomms11622
  10.1038/ncomms11622} {\path{doi:https://doi.org/10.1038/ncomms11622
  10.1038/ncomms11622}}.

\bibitem{wehrheimVI}
J.~Wehrheim, {Vortex invariants and toric manifolds. } (2008).
\newblock \href {http://arxiv.org/abs/0812.0299} {\path{arXiv:0812.0299}}.

\bibitem{baptistaDV}
J.~M. Baptista, N.~S. Manton, The dynamics of vortices on \({S}^2\) near the
  {B}radlow limit, J. Math. Phys. 44~(8) (2003) 3495--3508.
\newblock \href {https://doi.org/10.1063/1.1584526}
  {\path{doi:10.1063/1.1584526}}.

\bibitem{roldanQF}
A.~Rold\'an-Molina, M.~J. Santander, A.~S. Nunez, J.~Fern\'andez-Rossier,
  Quantum fluctuations stabilize skyrmion textures, Phys. Rev. B 92 (2015)
  245436.
\newblock \href {https://doi.org/10.1103/PhysRevB.92.245436}
  {\path{doi:10.1103/PhysRevB.92.245436}}.

\bibitem{psaroudakiQD}
C.~Psaroudaki, S.~Hoffman, J.~Klinovaja, D.~Loss, Quantum dynamics of skyrmions
  in chiral magnets, Phys. Rev. X 7 (2017) 041045.
\newblock \href {https://doi.org/10.1103/PhysRevX.7.041045}
  {\path{doi:10.1103/PhysRevX.7.041045}}.

\bibitem{atiyahMM}
M.~F. Atiyah, R.~Bott, The moment map and equivariant cohomology, Topology
  23~(1) (1984) 1--28.
\newblock \href {https://doi.org/10.1016/0040-9383(84)90021-1}
  {\path{doi:10.1016/0040-9383(84)90021-1}}.

\bibitem{cieliebakSV}
K.~Cieliebak, A.~R. Gaio, D.~A. Salamon, {$J$}-holomorphic curves, moment maps,
  and invariants of {H}amiltonian group actions, Internat. Math. Res. Notices
  10~(16) (2000) 831--882.
\newblock \href {https://doi.org/10.1155/S1073792800000453}
  {\path{doi:10.1155/S1073792800000453}}.

\end{thebibliography}

\end{document}